\documentclass[12pt, preprint]{aastex}

\def\d{{\rm d}}
\def\CK{{\cal K}}
\def\Rsun{{\rm R}_\odot}

\shorttitle{The structure of the solar core}
\shortauthors{Basu et al.}

\begin{document}

\title{Fresh insights on the structure of the solar core}

\author{Sarbani Basu}

\affil{Department of Astronomy, Yale University, P.O. Box 208101, New
Haven, CT 06520-8101; sarbani.basu@yale.edu}

\author {William~J.~Chaplin, Yvonne Elsworth} \affil{School of Physics
and Astronomy, University of Birmingham, Edgbaston, Birmingham B15
2TT, U.K.; w.j.chaplin@bham.ac.uk, y.p.elsworth@bham.ac.uk}

\author{Roger~New} \affil{Faculty of Arts, Computing, Engineering and
Sciences, Sheffield Hallam University, Sheffield S1 1WB, UK;
r.new@shu.ac.uk}

\author{Aldo~M.~Serenelli}

\affil{Max Planck Institute for Astrophysics
K. Schwarzschild Str. 1, Garching, D-85471, Germany;
aldos@mpa-garching.mpg.de}

\begin{abstract}

We present new results on the structure of the solar core, obtained
with new sets of frequencies of solar low-degree p modes obtained from
the BiSON network. We find that different methods used in extracting
the different sets of frequencies cause shifts in frequencies, but the
shifts are not large enough to affect solar structure results. We find
that the BiSON frequencies show that the solar sound speed in the core
is slightly larger than that inferred from data from MDI low-degree
modes, and the uncertainties on the inversion results are smaller.
Density results also change by a larger amount, and we find that solar
models now tend to show smaller differences in density compared to the
Sun.  The result is seen at all radii, a result of the fact that
conservation of mass implies that density differences in one region
have to cancel out density differences in others, since our models are
constructed to have the same mass as the Sun.  The uncertainties on
the density results are much smaller too.  We attribute the change in
results to having more, and lower frequency, low-degree mode
frequencies available. These modes provide greater sensitivity to
conditions in the core.

\end{abstract}


\section{Introduction}
\label{sec:intro}

Helioseismology is now an extremely successful and proven technique,
which can be used to infer the structure and dynamics of the interior
Sun (see. eg. Antia \& Basu 1994; Gough et al. 1995; Basu et
al. 2000a, b; Basu 2003; Couvidat et al. 2003; Schou et al. 1998a;
Pijpers 2006; Howe 2008, etc.). However, there is still considerable
uncertainty about the structure of the solar core.

Different data sets can often give results that show
discrepancies. Basu et al. (2000b) found that different solar
sound-speed results were obtained depending on whether data from the
Michelson Doppler Imager (MDI) or the ground-based Global Oscillation
Network Group (GONG) were used. The differences in the inferred
sound-speed in the solar core are illustrated in Figure~2 of Bahcall
et al.~(1998). The sensitivity of the inferred structure to the mode
sets has been noted in several investigations (see e.g. Basu et
al. 2000a, Turck-Chieze et al. 2001, etc.)

The model fitted to mode peaks in the oscillation power spectrum, as
part of the ``peak bagging'' procedures used to extract estimates of
the mode frequencies, can also give rise to discrepant inferences. For
example, Toutain et al. (1998) found that sound-speed differences in
the core were about 0.2\% to 0.3\% higher when they used frequencies
obtained by fitting asymmetric profiles to the asymmetric peaks in the
oscillation power spectrum, rather than what had been the usual
practice of fitting symmetric (Lorentzian) profiles. However, Basu et
al. (2000a) claimed that the differences were caused by a few
unreliable modes, and that the change in frequencies brought about by
fitting the more accurate asymmetric profiles would not be expected to
affect significantly the inversion results.

So, although the structure of the Sun, in particular the sound-speed
profile of the outer 80\% by radius, is known very well, the structure
of the core is relatively uncertain. The basic reason for this
uncertainty is that we only have frequency estimates of acoustic
modes, or p modes.  The p-modes that probe the inner 0.25$\Rsun$ of
the Sun are modes of low-degree $\ell$, i.e., with $\ell=0$, 1, 2 and
3.  All solar p-modes have their highest amplitudes close to the solar
surface, and hence even modes that probe the core (and there are
relatively few of them) are less sensitive to the conditions there
than in the outer parts of the solar interior.  Frequencies of solar
g-modes (buoyancy modes) would help reduce uncertainties, because they
have their highest amplitudes in the core. However, g modes are
extremely hard to detect, due to their small amplitudes at the solar
surface (see e.g., Appourchaux 2008; Appourchaux et al. 2000, 2006;
Gabriel et al. 2002; Elsworth et al. 2006).  There are recent
reports of the possible detection of the signature of several g modes
(see e.g., Garc\'{\i}a et al. 2008a, b). There are as yet no
convincing candidates for individual g-mode frequencies.  Thus, for
the moment, we still have to rely on p-modes to determine the
structure of the solar core.

The most reliable data on low-degree solar p-modes are obtained from
specialized instruments that observe the Sun as a star. The Birmingham
Solar-Oscillations Network (BiSON; Chaplin et al. 1996) is a network
of such instruments. Data from the network give precise estimates of
the frequencies of the low-degree p modes.  In this paper we use new
sets of data from BiSON to re-examine the structure of the solar core.

To extract information about the solar core properly through
inversions of solar frequencies we also need the higher,
intermediate-degree modes. Inversions are possible using only
low-degree modes, but the results have large errors and the resolution
can be poor (see e.g., Basu 2003). While this may be acceptable for
other stars, where for the foreseeable future the only modes we can
hope to observe are low-degree modes, this is not good enough for the
Sun.  Unfortunately instruments that determine frequencies of
low-degree modes precisely [such as BiSON or GOLF (Gabriel et
al. 1997)] cannot determine frequencies of intermediate-degree modes,
and instruments that determine frequencies of intermediate-degree
modes precisely (such as MDI and GONG) do not obtain as robust
frequencies at low degree as BiSON or GOLF.  The standard way to
overcome this problem is to combine precise low-degree data from one
source with precise intermediate-degree data from another. Combining
data from different sources requires caution since solar frequencies
are known to change with solar cycles (Elsworth et al. 1990; Libbrecht
\& Woodard 1990; Howe et al. 1999, etc).  It has been demonstrated by
Basu et al. (1996; 1997) that inverting combined sets when there is a
mismatch in the solar activity level of the low- and
intermediate-degree data sets can lead to misleading results about the
solar core. Thus one either has to use contemporaneous low- and
intermediate-degree data; or use a robust method for correcting one of
the sets of frequencies to the solar activity level of the other
set. Here, we compare results using both approaches.

The intermediate-degree mode frequencies that we use are those
estimated by Schou et al.~(1998b).  This set was obtained from
observations made by the MDI instrument on board SoHO during its first
year of operation and covers a period from May 9, 1996 to April 25,
1997. The activity level of the Sun was quite low over the entire
period that the data were collected, and hence frequency changes over
the year in question were not a matter of much concern. The BiSON data
sets we use in this paper are contemporaneous with the MDI set. In our
1997 paper (Basu et al. 1997), we had used BiSON data that were nearly
contemporaneous with the intermediate-degree data set obtained from
the LOWL instrument (Tomczyk et al. 1995). The BiSON data set used
then was a combination of frequencies obtained from an 8-month time
series, five 2-month time series, and low-frequency data obtained from
a 32-month time series. This combined set is referred to in Basu et
al. (1997) as the ``Best set'' and we use the same nomenclature here
to refer to this older set. Since 1997 there have of course been
improvements in our understanding of the properties of the solar
oscillation power spectrum and in the peak-bagging techniques used to
extract estimates of the frequencies. As a result, we believe it is
time to revisit the question of what the solar core looks like using
the BiSON data.

The rest of the paper is organized as follows: We describe our data
sets in \S~\ref{sec:data}. The inversion technique and reference
models are described in \S~\ref{sec:inv}. We present our results in
\S~\ref{sec:res}, and discuss their implications in \S~\ref{sec:disc}.

\section{Data}
\label{sec:data}

We use a number of different low-degree sets for this work, most of
them obtained with the BiSON instruments.  We have, however, also made
use of low-$\ell$ frequencies extracted from Sun-as-a-star
observations made by GOLF (Bertello et al. 2000, Garc\'ia et al. 2001)
and IRIS (Fossat et al. 2003), and from analysis of disc-integrated
MDI data (Toutain et al. 1998). The GOLF, IRIS and MDI frequencies are
in the literature. The BiSON frequencies come from our analysis of the
BiSON data, which is described below. Since the GOLF, IRIS and the
disc-integrated MDI data are not contemporaneous with the 360-day MDI
intermediate-degree set of Schou et al. (1998b) -- which hereafter we
shall refer to as data set MDI-1 -- for these sets we only use modes
with frequencies less than 1800$\mu$Hz. Solar-cycle related changes at
these low frequencies are negligible and smaller in size than the
uncertainties in the frequencies.  The data sets we have used for this
work, along with the names we use to identify the sets, are listed in
Table~\ref{tab:data}. All the BiSON sets are supplemented with $\ell
\ge 4$ data from MDI-1. The GOLF, MDIlow, and IRIS sets are
supplemented with MDI-1 data with $\nu > 1800\mu$Hz data for the
low-degree modes, and all the higher $\ell$ MDI-1 data.

The majority of the BiSON frequencies came from analysis of data
collected over two periods: a 360-day period commensurate with the
MDI-1 observations; and a much longer 4752-day period, beginning 1992
December 31 and ending 2006 January 3 (which includes the shorter
360-day period).  Estimates of the frequencies of very low-frequency
modes came from analysis of datasets prepared specifically to reduce
low-frequency background noise (see Appourchaux et al. 2000, and
Chaplin et al. 2002, for more details). These `low-frequency
optimized' datasets included one dataset prepared from observations
made over the 4752-d period; and two others prepared from shorter
2000-day and 3071-day periods within the 4752-day period.

There are two main advantages to using frequencies from the 4752-day
period and the other longer periods. First, the frequencies have
superior internal precision, e.g., frequency uncertainties in the
4752-day dataset are $\approx 3.5$-times smaller, on average, than
those in the 360-day dataset. Second, very low-frequency modes are
either more prominent, or only detectable, in the longer datasets.

The main downside to using BiSON frequencies from the longer periods
is that they do not cover the same levels of solar activity as the
360-day MDI period. The average activity over the 4752-day period, as
determined by various proxies of global activity, is about 60\,\%
higher than the average activity over the 360-day period.  Since an
increase in surface activity brings with it an increase in mode
frequencies, adjustments must be made to the BiSON frequencies to
correct them \emph{down} to values expected for the 360-day
period. Only then can the longer-period BiSON frequencies be usefully
combined with the 360-day MDI frequencies for inversion.

Solar-cycle variability is much less of a cause for concern in the
lower-frequency modes, since, as noted above, they show significantly
smaller solar-cycle frequency shifts than their higher-frequency
counterparts.  We therefore augmented the 360-day BiSON frequency sets
with frequencies of very low-frequency modes from the longer BiSON
datasets. The GOLF, MDI and IRIS frequency sets that we used were also
comprised only of low-frequency modes (having $\nu \le 1800\,\rm \mu
Hz$), and so the impact of the different epochs over which the various
data were collected should not have had a significant impact on the
modes that we used.

There is one other potential complication for combining the BiSON and
MDI frequencies. This complication arises from the fact that BiSON
observations are made of the ``Sun as a star''. The MDI observations are
in contrast made at high spatial resolution. The two types of
observation show a marked difference in sensitivity to modes having
certain combinations of $\ell$ and the azimuthal order $m$, and as a
result the frequencies may show a different response to the
azimuthally dependent surface activity.

We now go on to provide more detail on the BiSON analysis, and on the
important points raised above. We begin in \S~\ref{sec:bagging} with a
brief summary of how frequencies were estimated from the data. In
\S~\ref{sec:corr} we discuss how the long-dataset BiSON frequencies
were corrected to the activity level of the 360-day MDI period. Then
in \S~\ref{sec:mismatch} we explain the inherent mismatch between the
BiSON Sun-as-a-star and MDI resolved-Sun data, and the extent to which
it is an issue for the frequency data used in our studies, and 
in \S~\ref{subsec:comp} we compare the different frequency sets.

\subsection{Extraction of mode frequencies from BiSON data}
\label{sec:bagging}

A set of BiSON Sun-as-a-star observations gives a single time series
whose power frequency spectrum contains many closely spaced mode
peaks.  Parameter estimation must contend with the fact that within
the non-radial ($\ell > 0$) mode multiplets the various $m$ lie in
very close frequency proximity to one another. Suitable models, which
seek to describe the characteristics of the $m$ present, must
therefore be fitted to the components simultaneously. Furthermore,
overlap between modes adjacent in frequency means the modes are
usually fitted in pairs.

We fitted multi-component models to modes in power frequency spectra
of the various BiSON datasets to extract estimates of the mode
frequencies.  This fitting was accomplished by maximizing a likelihood
function commensurate with the $\chi^2$, 2-d.o.f. statistics of the
power spectral density.  We adopted the usual approach to fitting the
Sun-as-a-star spectra, and the low-$\ell$ modes were fitted in pairs
($\ell=0$ with 2, and $\ell=1$ with 3). Further background on details
of the procedures may be found in Chaplin et al. (1999).

\subsection{Correction of long dataset BiSON frequencies to 360-day
  MDI period}
\label{sec:corr}

Our procedure for removing the solar-cycle frequency shifts rests on
the assumption that variations in certain global solar activity
indices may be used as a proxy for the low-$\ell$ frequency shifts,
$\delta\nu_{n\ell}(t)$. We assume the correction can be parameterized as a
linear function of the chosen activity measure, $A(t)$. When the
10.7-cm radio flux (Tapping \& De~Tracey 1990) is chosen as the proxy,
this assumption is found to be reasonably robust (e.g., Chaplin et
al. 2004a) at the level of precision of the data.

Consider then the set of measured BiSON mode frequencies,
$\nu_{n\ell}(t)$, that we wish to `correct'. Let us take the example
where the data come from observations collected over the $t=4752\,\rm
d$ period, when the mean level of the 10.7-cm radio flux was $\left<
A(t) \right> = 121 \times 10^{-22}\,\rm W\,m^{-2}\,Hz^{-1}$. Our
intention it to correct these frequencies to the mean level of $\left<
A(t) \right>_{360} = 75 \times 10^{-22}\,\rm W\,m^{-2}\,Hz^{-1}$
observed over the 360-day period covered by the MDI data. The mean
activity in this 360-day period only just exceeds the canonical
quiet-Sun level of the radio flux, which, from historical observations
of the index, is usually fixed at $64 \times 10^{-22}\,\rm
W\,m^{-2}\,Hz^{-1}$ (see Tapping \& DeTracey 1990).  The magnitude of
the solar-cycle correction -- which must be subtracted from the raw
frequencies -- will then be:
 \begin{equation}
 \delta \nu_{n\ell}(t) = g_\ell \cdot {\cal F}[\nu] 
                   \cdot [\left< A(t) \right> - \left< A(t) \right>_{360}].
 \label{eq:corr}
 \end{equation}
The $g_{\ell}$ are $\ell$-dependent factors that calibrate the size of
the shift. These factors are required because the Sun-as-a-star shifts
alter significantly with $\ell$ (see \S~\ref{sec:mismatch} below).  To
determine the $g_\ell$, we divided the 4752-day timeseries into 44
independent 108-day segments. The resulting ensemble was then
analyzed, in the manner described by Chaplin et al. (2004a), to
uncover the dependence of the solar-cycle frequency shifts on the
10.7-cm radio flux. The ${\cal F}[\nu]$ in Equation~\ref{eq:corr} is a
function that allows for the dependence of the shift on mode
frequency. Here, we used the determination of ${\cal F}[\nu]$ to be
found in Chaplin et al. (2004a, b).

Uncertainty in the correction is dominated by the errors on the
$g_\ell$. These errors must be propagated, together with the formal
uncertainties from the mode fitting procedure, to give uncertainties
on the corrected frequencies, $\nu_{n\ell}(t)-\delta
\nu_{n\ell}(t)$. The corrected uncertainties are, on average, about
10\,\% larger than those in the raw, fitted frequencies.

At frequencies below about $\approx 1800\,\rm \mu Hz$, estimated
corrections for the 4752-day dataset, and the other long datasets,
were smaller in size than the fitted frequency uncertainties. We
therefore felt justified in augmenting the 360-day BiSON frequency set
with very low-frequency estimates from the longer BiSON datasets.

\subsection{On the inherent mismatch between Sun-as-a-star and
  resolved-Sun frequencies}
\label{sec:mismatch}

Extant Sun-as-a-star observations, such as those of the BiSON, are
made from a perspective in which the plane of the rotation axis of the
Sun is nearly perpendicular to the line-of-sight direction.  This
means that only components with $\ell+m$ even have non-negligible
visibility. Resolved-Sun observations -- like MDI -- in contrast have
good sensitivity to all components in detected modes. Sun-as-a-star
and resolved-Sun frequency data must as a result be combined very
carefully since there there can be an inherent, underlying mismatch
between frequency determinations from the two types of instrument, as
will be explained below.

The resolved-Sun data allow for a direct measurement of the frequency
centroids of the non-radial modes, because all components are
detectable. The frequency centroids carry information on the
spherically symmetric component of the internal structure, and are the
input data that are required for the hydrostatic structure inversions.
In the case of the Sun-as-a-star data, some components are missing,
and the centroids must be estimated from the subset of visible
components. In the complete absence of the near-surface activity, the
$\ell+m$ odd components `missing' in the Sun-a-as-a-star data would be
an irrelevance. All mode components would be arranged symmetrically in
frequency, meaning centroids could be estimated accurately from the
subset of visible components. A near-symmetric arrangement is found at
the epochs of the modern cycle minima. However, when the observations
span a period having medium to high levels of activity -- as a long
dataset by necessity does -- the arrangement of components is no
longer symmetric. The frequencies given by fitting the Sun-as-a-star
data then differ from the true centroids by an amount that is
sensitive to $\ell$. The $\ell$ dependence arises because in the
Sun-as-a-star data modes of different $\ell$ comprise visible
components having different combinations of $\ell$ and $|m|$; and
these different combinations show different responses (in amplitude
and phase) to the spatially non-homogeneous surface activity.

Is it possible to correct the Sun-a-as-a-star frequencies to remove
the mismatch? The answer is yes, and the procedure requires knowledge
of the strength and spatial distribution of the surface activity over
the epoch in question. Chaplin et al. (2004c) and Appourchaux \&
Chaplin (2007) show how to make the correction, using the so-called
even $a$ coefficients from fits for the resolved-Sun frequencies.

Does the mismatch matter for the datasets used here? We should not
expect the mismatch to be a cause for concern between the 360-day
BiSON frequencies and 360-day MDI frequencies. That is because the
surface activity is low throughout this period. We have verified that
application of the correction procedure outlined in Appourchaux \&
Chaplin (2007) does not alter significantly the 360-day BiSON
frequencies, nor does it affect significantly the results of the
structure inversions.

There will be a more serious mismatch between the 4752-day BiSON
frequencies and the 360-day MDI frequencies. However, the solar-cycle
frequency correction procedure, outlined in \S~\ref{sec:corr} above,
adjusts \emph{by definition} the BiSON frequencies to values expected
at low levels of activity; as such this in principle removes the
Sun-as-a-star and resolved-Sun mismatch without the need to apply a
further correction.

\subsection{Comparison of the frequencies}
\label{subsec:comp}

\begin{figure}
\epsscale{0.7}
\plotone{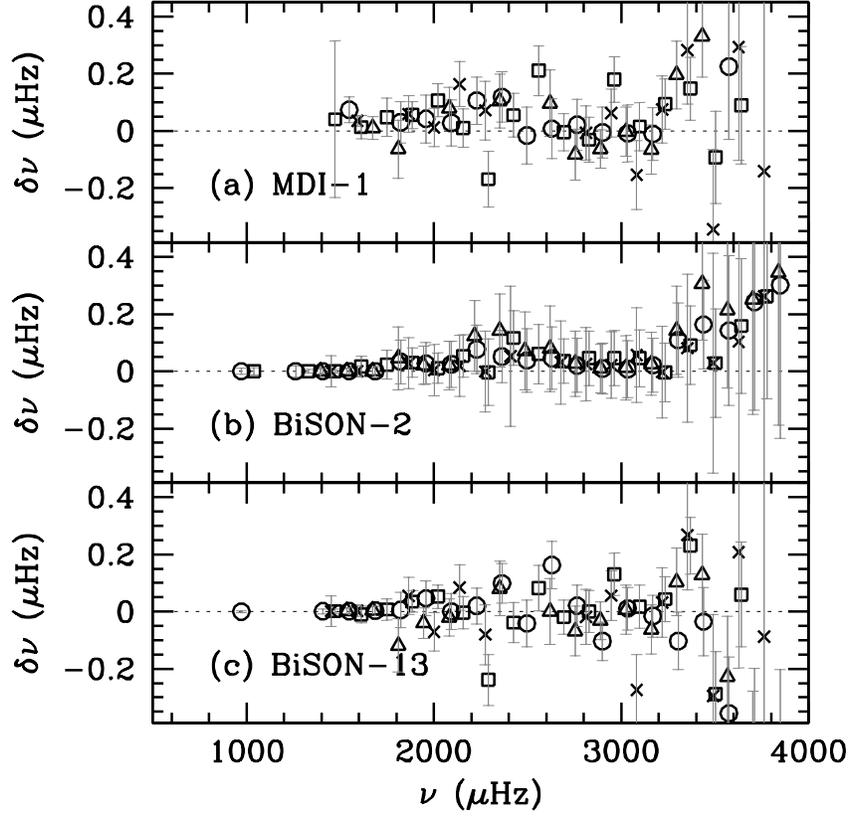}

\caption{Frequency differences between the BiSON-1 set and: (a) the
MDI-1 set; (b) the BiSON-2 set; (c) the BiSON-13 set. The differences are in the sense
(BiSON-1$-$ other set.  Circles show $\ell=0$, squares show $\ell=1$,
triangles show $\ell=2$ and crosses show $\ell=3$ modes.}

\label{fig:freqdif}
\end{figure}

Panel (a) of Figure~\ref{fig:freqdif} shows frequency differences
between the BiSON-1 set and the MDI-1 set, while panel (b) of the same
figure shows differences between the BiSON-1 set and the BiSON-2 set.
The differences in panels (a) and (b) both have the same basic
structure. This structure comes from the fact that the BiSON-2 and
MDI-1 sets were derived from fits to the oscillation power spectra
that used a symmetric (Lorentzian) profile for each mode peak, while the
BiSON-1 set came from fits made using a more accurate asymmetric
profile. There is also likely to be a contribution to the BiSON-1 minus
MDI-1 residuals from differences in the peak-bagging methodologies
applied to Sun-as-a-star (BiSON) and resolved-Sun (MDI) data.

The MDI-1 mode set has fewer low-degree modes than the BiSON-1 set,
and as a result we would expect to see differences between inversion
results obtained by the MDI-1 set and those obtained with BiSON sets
supplemented by $\ell\ge4$ modes from the MDI-1 set. In order to judge
the effect of the number of low-degree modes in the set, we have also
inverted data using only those modes of the BiSON-1 set that are also
present in the MDI-1 set. We call this restricted set the BiSON-1m set
for ease of reference.

Differences between the BiSON-1 and 4752-day BiSON-13 set, shown in
panel (c) of Figure~\ref{fig:freqdif}, are on the whole very small
because both sets came from asymmetric-profile fits. There are a few
$>1\sigma$ outliers, which probably reflect the impact on the fitting
of different realization noise in the two sets of data. The observed
differences do also increase in size to above $1\sigma$ at high
frequencies. This is the part of the oscillation spectrum where the
mode peaks are very wide (the modes are heavily damped), meaning the
peaks of mode components, and modes, adjacent in frequency begin to
overlap. Some modest bias in the estimation of the frequencies can
result, and this will be more severe in the shorter BiSON-1 set than
in the longer BiSON-13 set because of the inferior frequency
resolution of the 360-day BiSON-1 data.

\begin{figure*}
\epsscale{1.0}
 \plotone{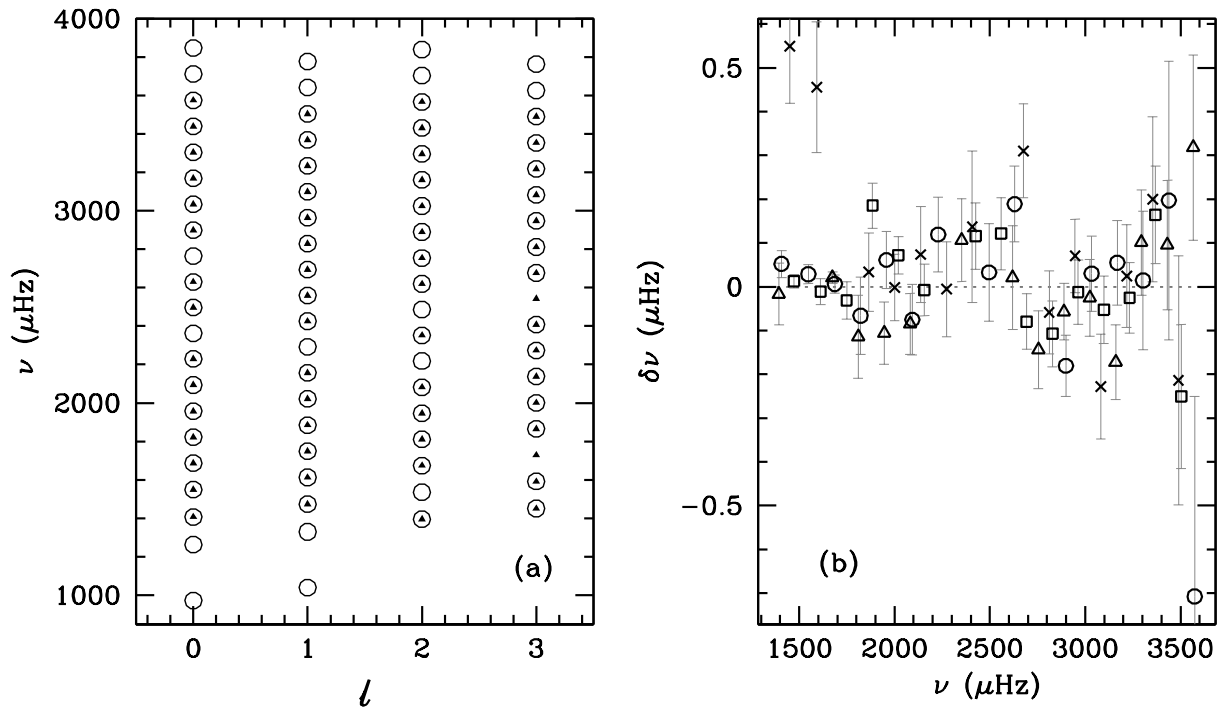}

 \caption{(a) The low-degree $\ell-\nu$ diagram for the BiSON-1 set
(circles) and the ``Best set'' of Basu et al. (1997; triangles). (b)
The frequency difference between the BiSON-1 set and the ``Best
set''. Circles are for $\ell=0$, squares for $\ell=1$, triangles for
$\ell=2$ and crosses for $\ell=3$ modes. The differences are in the
sense BiSON-1$-$''Best set''.}

 \label{fig:best}
 \end{figure*}

Panel (a) of Figure~\ref{fig:best} shows frequency differences between
the BiSON-1 and the 1997 ``Best set''. As can be seen in the figure,
the new BiSON-1 set has filled up some of the gaps in the $\ell-n$
space. Also, the new set extends to somewhat lower as well as higher
frequencies.  The frequency differences shown in panel (b) of the same
figure have a definite structure as a function of frequency, which has
a similar pattern to the structure in panels (a) and (b) of
Figure~\ref{fig:freqdif}. These differences are again largely due to
the fact that the ``Best set'' was derived from fits that used a
symmetric profile for the mode peaks. It should be noted that some of
the difference in the frequencies probably also comes from other
improvements to the peak-bagging routines that have been implemented
since the mid 1990s.

Since the overall differences are not random, we might again expect to
find some differences in the inferred structure of the solar core

\section{The inversion technique}
\label{sec:inv}

Inversion for solar structure is complicated because the problem is
inherently non-linear. The inversion generally proceeds through a
linearization of the equations of stellar oscillations, using their
variational formulation, around a known reference model (see e.g.,
Dziembowski et al.~1990; D\"appen et al.~1991; Antia \& Basu 1994;
Dziembowski et al.~1994, etc.).  The differences between the structure
of the Sun and the reference model are then related to the differences
in the frequencies of the Sun and the model by kernels.  Nonadiabatic
effects and other errors in modeling the surface layers give rise to
frequency shifts which are not accounted for by the variational
principle.  In the absence of any reliable formulation, these effects
have been taken into account by including an arbitrary function of
frequency in the variational formulation (e.g., Dziembowski et
al.~1990).

The fractional change in frequency of a mode can be expressed in terms
of fractional changes in the structure of model characteristics, for
example, the adiabatic sound speed $c$ and density $\rho$, and a
surface term.  The frequency differences can be written in the form
(e.g., Dziembowski et al.~1990):
 \begin{equation}
 {\delta \nu_i \over \nu_i}
 = \int_0^{R_\odot} K_{c^2,\rho}^i(r){ \delta c^2(r) \over c^2(r)}\; {\rm d} r +
  \int_0^{R_\odot} K_{\rho,c^2}^i(r) {\delta \rho(r)\over \rho(r)}\; {\rm d} r
  +{F_{\rm surf}(\nu_i)\over I_i}
 \label{eq:inv}
 \end{equation}
Here $\delta \nu_i$ is the difference in the frequency $\nu_i$ of the
$i$th mode between the data and the reference model, where $i$
represents the pair ($n,\ell$), $n$ being the radial order and $\ell$
the degree.  The kernels $K_{c^2, \rho}^i$ and $K_{\rho, c^2}^i$ are
known functions that relate the changes in frequency to the changes in
the squared sound speed $c^2$ and density $\rho$ respectively, and
$I_i$ is the mode inertia.  The kernels for the $(c^2,\rho)$
combination can be easily converted to kernels for others pairs of
variables like $(\rho,\Gamma_1)$, with no extra assumptions (Gough
1993).  The term $F_{\rm surf}$ is the ``surface term'', and takes
into account the near-surface errors in modeling the structure.

Equation~(\ref{eq:inv}) constitutes the inverse problem that must be
solved to infer the differences in structure between the Sun and the
reference solar model.  We carried out the inversions using the
Subtractive Optimally Localized Averages (SOLA) technique (Pijpers \&
Thompson 1992; 1994). The principle of the inversion technique is to
form linear combinations of equation~\ref{eq:inv} with weights
$w_i(r_0)$ chosen such as to obtain an average of $\delta c^2/c^2$
localized near $r = r_0$ (the `target radius') while suppressing the
contributions from $\delta \rho/\rho$ and the near-surface errors.  In
addition, the statistical errors in the combination must be
constrained.  If successful, the result may be expressed as
 \begin{equation}
 \int \CK(r_0, r) {\delta f_1(r) \over f_1(r)} \d r
 \simeq \sum w_i(r_0) {\delta \omega_i \over \omega_i} \; ,
 \label{eq:avker}
 \end{equation}
where the $\CK(r_0, r)$, the averaging kernel at $r=r_0$, is defined
as
\begin{equation}
 \CK(r_0, r) = \sum w_i(r_0) K_{1,2}^i(r)  \; ,
\label{eq:avdef}
\end{equation}
of unit integral, and determines the extent to which we have achieved
a localized measure of $\delta c^2/c^2$.  In particular, the width in
$r$ of $\CK(r_0,r)$, here calculated as the distance between the first
and third quartile point, provides a measure of the
resolution. Although we attempt to obtain averaging kernels at the
target radius $r_0$ this is not always possible, and in the core very
often the kernel is localized at a different point. This point, unless
the averaging kernel is really non-localized, coincides with the
second quartile point of the averaging kernel. As a result all our
results are plotted as a function of the 2nd quartile points of the
averaging kernels obtained at the end of the inversion process.
Details of how SOLA inversions are carried out and how various
parameters of the inversion are selected were given by Rabello-Soares
et al. (1999).

Since the solar core is difficult to invert for, we also inverted the
data using the Regularized Least Squares (RLS) technique.  Given the
complementary nature of the RLS and SOLA inversions (see Sekii 1997
for a discussion), we can be more confident of the results if the two
inversions agree. Details on RLS inversions and parameter selections
can be found in Antia \& Basu (1994) and Basu \& Thompson (1996).  In
this work we invert for the relative difference of the squared sound
speed, i.e.  $\delta c^2/c^2$ between the Sun and the reference solar
model using $(c^2,\rho)$ kernel pairs, and for the relative density
difference $\delta\rho/\rho$ between the Sun and reference model using
$(\rho,\Gamma_1)$ kernel pairs.

Our main reference model is model BP04 of Bahcall et al. (2005).  We
also use two other models in the discussion of our results, model S of
Christensen-Dalsgaard et al.~(1996), and model BSB(GS98) of Bahcall et
al. (2006), which in this paper we refer to as simply model BSB. All
the models are standard solar models, but have been constructed with
somewhat difference input physics. The characteristics and physics
inputs of the models are listed in Table~\ref{tab:models}.

Most of the structure inversion results found in the literature have
been obtained with model S as the reference model. However, the input
physics in model S is now somewhat outdated, in particular the
equation of state is known to have problems under the conditions found
in the solar core. The opacities are also somewhat outdated. As a
result we use BP04 as our reference model.  Model BP04 is very similar
to model S, but it is constructed with an improved equation of state
and newer opacities. Model BSB is in contrast quite different from the
other two models. In particular it uses OP opacities (Badnell et
al. 2005) instead of OPAL opacities (Iglesias \& Rogers 1996). The OP
opacities are somewhat smaller than OPAL opacities for conditions
expected in the solar core, but somewhat larger than OPAL opacities at
the base of the convection zone. The core structure has also been
affected by the new and lower $^{14}$N($p$,$\gamma$)$^{15}$O reaction
rate (Formicola et al.  2004).

It should be noted that all the models that we use are high-$Z/X$ models,
with $Z/X$ values adopted from either Grevesse \& Noels (1993) or  from
Grevesse \& Sauval (1998). Asplund et al. (2005) compiled a table of solar 
abundance determinations based on new techniques that showed that solar 
metallicity is about 30\% lower than previous estimates. This led to large
changes in the structure of standard solar models, and in particular, made the
agreement between the Sun and solar models much worse [see
Basu \& Antia (2008) for a review of the problem]. Since the aim of this paper
is to take a fresh look into the structure of the Sun rather than study solar
models, we restrict ourselves to models that are known to agree well 
with the Sun.

\section{Results}
\label{sec:res}

\subsection{Sound speed}
\label{subsec:sound}

The relative differences in the squared sound speed between the Sun
and model BP04, as obtained with the different BiSON sets, are plotted
in Figure~\ref{fig:csqbison}.  Also plotted is a close-up of the
differences in the core. We can see from the figure that the 1-year
BiSON sets combined with $\ell \ge 4$ modes from the MDI 360 day set
give very similar results, but that the results are different from
those obtained with the MDI-1 set, which is the MDI 360-day set for
all degrees. The differences between the MDI-1 and BiSON sets are seen
below about $0.35R_\odot$, where the $\ell=3$ modes begin to influence
the results. While we were able to get smaller errors in the inversion
results, we were not able to push the inversions much deeper without
increasing the errors.

The differences between the MDI and BiSON results are not merely a
result of having more modes. If that were so, the result of inverting
set BiSON-1m would have been similar to the MDI-1 result. While this
is indeed true between about $0.2$ and $0.3R_\odot$, in deeper layers
the BiSON-1m set gives very similar results to the other BiSON sets,
pointing to the fact that the frequencies themselves, as well as the
lower errors of the BiSON frequencies, play a r\^ole. 

The sound-speed obtained with the symmetric BiSON-2 set is marginally
higher than that obtained with the asymmetric BiSON-1 set, however,
the results agree within errors.  Since all the 1-year BiSON sets give
similar results, and since peaks in the solar oscillation power
spectrum are known to be asymmetric, in further discussion of the
360-d BiSON inversions we shall use the results from BiSON-1, not
BiSON-2.

As far as the BiSON-13 set is concerned, it lets us go marginally
deeper into the core, with the innermost averaging kernel centered at
0.062$\Rsun$. We get lower sound speeds than BiSON-1 but higher speeds
than MDI-1 in the deepest regions, as is seen in
Figure~\ref{fig:csq13}. The BiSON-13 results above 0.2 $R_\odot$ lie
between the BiSON-1 and MDI results; while between 0.1 and 0.2
$R_\odot$ they lie slightly above the BiSON-1 and MDI-1 results.  {
Figure~\ref{fig:avkcsq} shows the innermost averaging kernel, as well
as two others obtained with the BiSON-13 and BiSON-1 sets.  Also shown
are the ``cross-term'' kernels corresponding to the innermost
averaging kernels.  Note that the peak of the BiSON-13 averaging
kernel is closer to the centre.  }

Figure~\ref{fig:csqbest} shows the sound-speed difference obtained by
the BiSON-1 set compared with that of the ``Best set'' of Basu et
al. (1997). We see that the differences in structure are small. With
the new set we can though push closer to the core, which we believe is
a result of the lower-frequency modes in the new set. One point should
be noted: in the Basu et al.~(1997) paper, we had plotted the results
against the ``target'' radius, instead of the 2nd quartile point as we
have done here. Our experience with inversions in the intervening
period leads us to believe that plotting the results against the 2nd
quartile point is more correct, since they are the points where the
averaging kernels are localized.

The BiSON-1 and MDI-1 results are compared with the other data sets
listed in Table~\ref{tab:data} in Figure~\ref{fig:csq_comp}. For the
external data sets, we show two inversions, one marked ``All'' and the
other marked ``Weeded.'' The inversions marked ``All'' used all the
$\nu < 1800\mu$Hz modes in the sets, while the ones marked ``Weeded''
had a few modes removed, usually because they had large residuals in
the RLS inversions.  The MDIlow ``All'' results show a steeper rise in
solar sound-speed near the core. However, RLS inversion of this set
show that two modes ($\ell=2,n=6$ and $\ell=2,n=7$) had extremely
large residuals (16 and 22$\sigma$ respectively).  This problem had
also been noted earlier by Basu et al.~(2000a), who used different
reference models. The weeded results are quite similar to the BiSON
results. The IRISlow set was problematic in that many modes had high
residuals. The worst offenders were the $\ell=1,n=7$, $\ell=1,n=10$
and $\ell=2,n=7$ modes. With these modes in place, the IRISlow data
implied that the solar sound-speed increased enormously compared to
the reference model at radii below $0.1R_\odot$. With these modes
removed the results are still higher than the BiSON results, but
within 1$\sigma$.

The results for the two GOLF sets are much more interesting. The
unweeded results (``All'') show a marked departure from the BiSON and
MDI results below 0.15$R_\odot$.  These two GOLF sets contain very
low-frequency detections that are controversial and extremely
uncertain, which have not been verified independently by other data.
The GOLF1low set set has an $\ell=0,n=5$ mode at $825.202\pm
0.005\mu$Hz and an $\ell=0,n=3$ mode at $535.743\pm 0.003\mu$Hz. The
GOLF2low set has an $\ell=0,n=3$ at $535.729\pm0.009\mu$Hz. The lowest
BiSON-1 frequency, for comparison, is the $\ell=0,n=6$ mode at
$972.613\pm0.002\mu$Hz. We note that this mode has been observed in
BiSON and GOLF data.

The $\ell=0,n=3$ mode changes the inversion results for the GOLF2low
set completely. Given that there are no other modes in the set at
comparably low frequencies, this raises a serious question mark over
whether this is a real effect. In the absence of this mode, the
GOLF2low results (``Weeded'' in Figure~\ref{fig:csq_comp}) match the
BiSON-1 results almost exactly. The situation is the same with the
GOLF1low results. { Note that the same inversion parameters have
been used for both the ``Weeded'' and ``All'' sets.} The $\ell=0,n=3$
mode appears to exert a disproportionate influence, the $\ell=0,n=5$
less so, but it also pulls down the solution. Without these two modes,
the sound-speed difference obtained is again very similar to that of
the BiSON-1 set. We discuss the issue of low-frequency modes further
in \S~\ref{subsec:den}. Thus, if for the moment we ignore the effect
of the $\ell=3$, $n=3,5$ modes, it appears that the solar sound-speed
still shows a dip in the sound-speed difference with respect to the
solar model around $0.2R_\odot$ and rises to positive values at radii
below $0.1R_\odot$. The effect of the two very low-frequency modes in
lowering the estimated sound speed in the core has also been noted by
Turck-Chieze et al.~(2001). { The changes between the ``All'' and
``Weeded'' results are caused by differences in the inversion
coefficients. For SOLA inversions, the modes that were weeded out
changed the inversion coefficients of the neighbouring low-degree
modes by a large amount. The changes were much smaller in the case of
RLS inversions, and because the low-degree modes have large residuals,
the changes did not affect the solution.}

\subsection{Density}
\label{subsec:den}

The density differences between the Sun and model BP04, as obtained by
the different BiSON sets, are shown in
Figure~\ref{fig:bisondensity}(a). Also shown in the figure is the
result of inverting the MDI-1 set. As can be seen from the figure, the
BiSON sets give fairly similar results, within errors. However, the
MDI results are different at all radii, not just close to the core. At
first glance this is surprising since the only difference between the
MDI and BiSON sets are the $\ell < 4$ modes, and one would expect
differences only in the core, as was found for the sound-speed
difference.  However, the result is not really surprising.
Density inversions are carried out by imposing the condition that the
mass of the reference model is the same as that of the Sun. Thus the
density differences must integrate out to zero, i.e.,
\begin{equation}
\int 4\pi r^2\delta\rho=\int 4\pi r^2\rho {\delta\rho\over\rho}=0.
\label{eq:deninv}
\end{equation}
The density $\rho$ in the core is large, and thus a small relative
difference $\delta\rho/\rho$ in the core will lead to a large
difference in the regions where the density is small.  The higher
number of low-degree models in the BiSON sets allows us determine the
core structure much better than with the MDI set: we find a smaller
difference in the core, which implies a smaller difference at all
radii. That said, the difference between the BiSON and MDI results is
not merely a matter of the larger BiSON mode set. If we only use BiSON
frequencies of those modes that are present in the MDI set (i.e., set
BiSON-1m), the result lies between the MDI and the other BiSON sets. This
can be seen in Figure~\ref{fig:bisondensity}(a). These results show us
that when we attempt to infer the solar density it is not enough to have
good intermediate and high-degree modes: it is more important to have
reliable data on low-degree modes. Given that the BiSON low-degree
sets still have gaps and do not yet have very low-frequency modes, we
can be sure that solar density results will change if we get better
low-degree sets.

In Figure~\ref{fig:bisondensity}(b) we show the relative differences
in the inferred solar density obtained from the different sets. The
differences are shown with respect to the results obtained with the
BiSON-1 set. As can be seen, the differences are completely
systematic, even when within errors as is the case for
BiSON-2. Judging by the difference between the results of the MDI-1
set and the results of the BiSON sets, we estimate that despite the
small statistical errors, solar density inversions are uncertain at a
level of about 0.6\%.  Figure~\ref{fig:rho13} shows the density
differences obtained with the BiSON-13 set compared to those obtained
with the BiSON-1 and MDI sets. The BiSON-13 set seems to imply a
low-density solar core.

Figure~\ref{fig:rho_best} compares the density differences obtained
using the ``Best set'' of Basu et al. (1997) and the BiSON-1 set. Note
that the density differences obtained with the ``Best set'' are larger
in general, and in the core they imply that the density in the Sun
appears to be much greater than in the reference model.  This result
is in seeming contradiction to the result reported by Basu et
al.~(1997) (even though that was for a different reference model, the model S),
which was that the solar density was found to be much lower than the
density in the reference model.

This discrepancy appears to be a result of the use of the $(\rho,Y)$
kernel combination by Basu et al. (1997).  Density difference results
obtained for the ``Best set'' using model S as the reference and with
$(\rho,\Gamma_1)$ and $(\rho, Y)$ kernels are shown in
Figure~\ref{fig:rho_yker}. With $(\rho,\Gamma_1)$ kernels we see that
the inferred density difference between the Sun and the model is
positive in the core. Transformation to kernels involving $Y$ requires
the assumption that the equation of state and the heavy element
abundance are known exactly. Basu \& Christensen-Dalsgaard (1997)
showed that the density results obtained with these $(\rho, Y)$
kernels are sensitive to errors in the equation of state assumed for
the reference model. Model S was constructed with an equation of state
that was found to be deficient for conditions found in the solar core
(Elliott \& Kosovichev 1998), and as a result we believe that the
results obtained using the $(\rho, \Gamma_1)$ variable combination are
more robust than those obtained with $(\rho, Y)$, even though it is at
the expense of larger statistical errors. Basu et al. (1997) had
discussed the possibility of systematic errors in the density results
caused by the particular choice of kernels, but had not tested the
results obtained using other sets of kernels.

Figure~\ref{fig:rhoall} shows the density differences obtained for the
four external sets --- MDILow, IRISLow, GOLF1Low, and GOLF2low. We
only show the results of the weeded sets. Although there are
differences in the results obtained by the various sets, only set
IRISlow gives completely discrepant results, probably implying that
some of the frequencies are not accurate. While all other differences
lie within a few $\sigma$ there are systematic trends as expected from
the conservation of mass constraint (see above). The two GOLF results
are again interesting. Although the results obtained from the weeded
sets shown in Figure~\ref{fig:rhoall} are very similar to those
obtained using the BiSON sets, the unweeded results are
troublesome. Figure~\ref{fig:golf} shows both the weeded and unweeded
results obtained with the GOLF sets. We show both SOLA and RLS
results. The striking feature of the unweeded results is that the SOLA
results do not match the RLS results at all. This again leads us to
believe that there could be some problems with the very low-frequency
$\ell=0$ modes that were weeded out from the two sets. One of the
drawbacks of the SOLA method is the implicit assumption that the
frequency differences and the errors associated with them are
correct. However, sometimes the observational errors can be either
under- or over-estimated, or the modes may have unusual
characteristics (e.g., much smaller line widths than modes adjacent in
frequency), in which case the inversion results can be misleading.
RLS inversions are less prone to be affected by such outliers, and
hence we are more inclined to believe the RLS results, which agree
with the weeded SOLA results.

\section{Discussion}
\label{sec:disc}

We have used a number of BiSON low-$\ell$ data sets, combined with
$\ell > 3$ data from the MDI 360-day set, to obtain sound speed and
density differences between our reference model BP04 and the Sun. We
find that we are able to go slightly deeper into the core than we had
been able to do with earlier BiSON sets. While the sound-speed results
are similar to what we had seen earlier (e.g., Basu et al. 1997), the
density results show larger differences.  Since the new sets of BiSON
frequencies have more modes, and smaller estimated uncertainties, we
believe that our current results are more reliable.

Differences in inversion results obtained by using as input
frequencies estimated from symmetric and asymmetric fits to the
oscillation power spectra are within errors. This is what had also
been found for GOLF data by Basu et al. (2000a).  Similarly the
inversion results obtained from frequencies with and without
Sun-as-a-star corrections are very similar. The differences in density
results are larger, but although they are systematic, the differences
are still within errors. The BiSON-13 set, which has low frequency
data from a 4752-day long spectrum, allows us to get results with
somewhat smaller errors.  The solar sound-speed and density results
obtained with this set are tabulated in Table~\ref{tab:seismic}.

\begin{figure*}
\epsscale{1.0}
\plotone{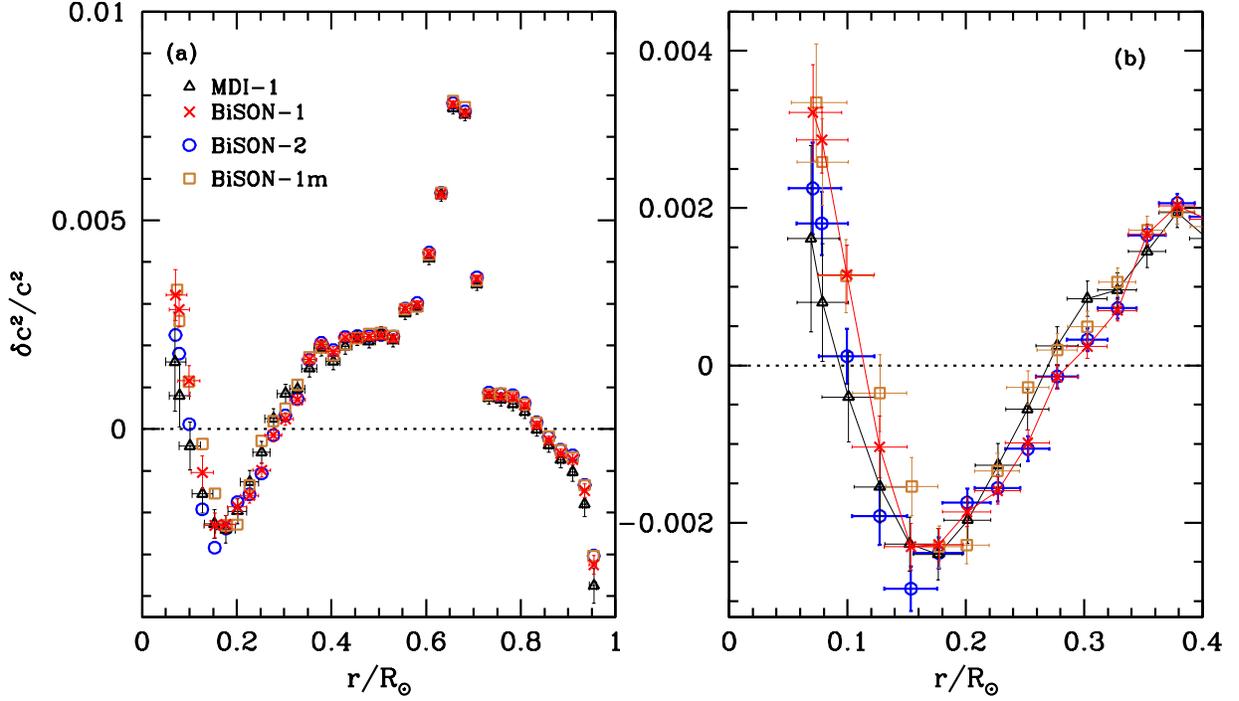}

\caption{The relative difference in the squared sound-speed between
the Sun and reference model BP04 obtained by inverting the different
data sets marked in the figure. SOLA inversion results are shown. Panel (a) shows the entire radius
range, while panel (b) focuses on the core. The vertical error bars
are a measure of the errors in the inversion and come from the
uncertainties in the frequencies propagated through the inversion
process.  The horizontal error bars are a measure of the resolution of
the inversions, and are the distance distance between the first and
third quartile points of the averaging kernels obtained from the
inversions. Only two sets of errors are shown in panel (a) for the
sale of clarity. In panel (b) we have joined the points corresponding
to MDI-1 and BiSON-1 sets to guide the eye.}

\label{fig:csqbison}
\end{figure*}

\begin{figure}
\epsscale{0.8}
\plotone{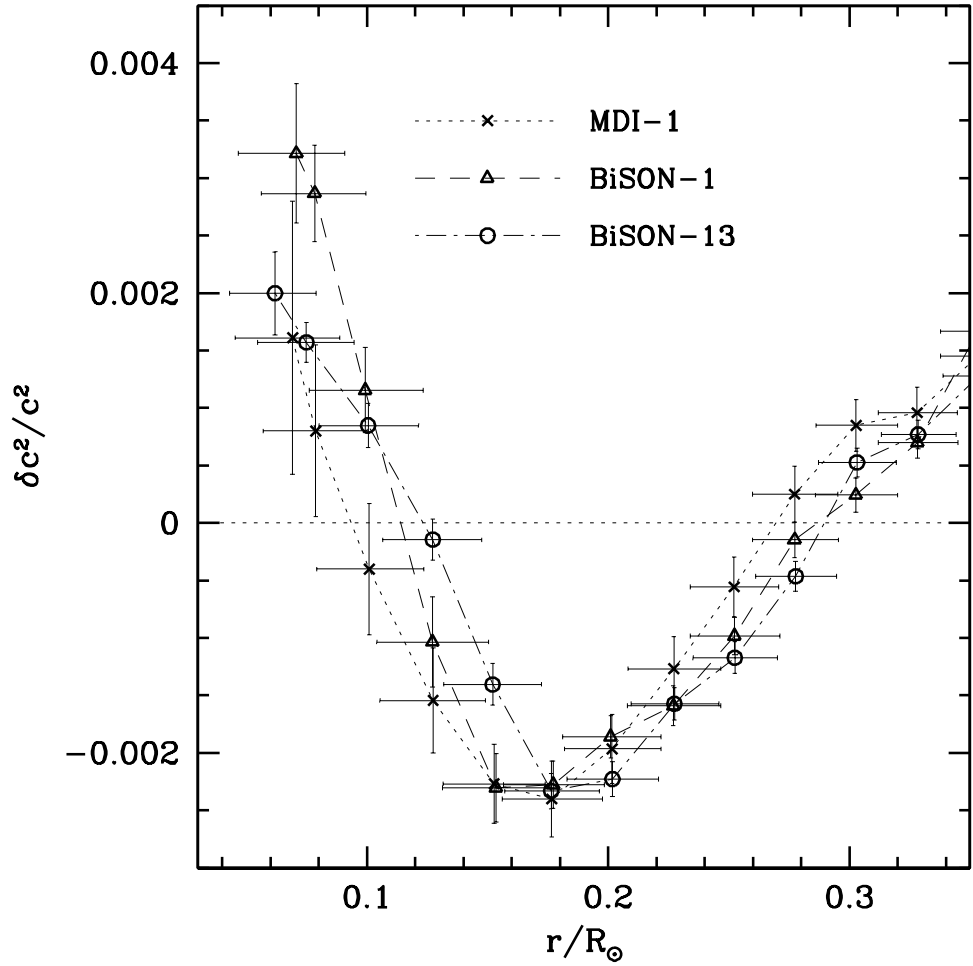}

\caption{The relative difference in the squared sound-speed between
the Sun and reference model BP04 obtained by inverting set
BiSON-13. SOLA inversion results are shown. For comparison, we also show results from sets MDI-1 and
BiSON-1.}

\label{fig:csq13}
\end{figure}

\begin{figure}
\epsscale{0.8}
\plotone{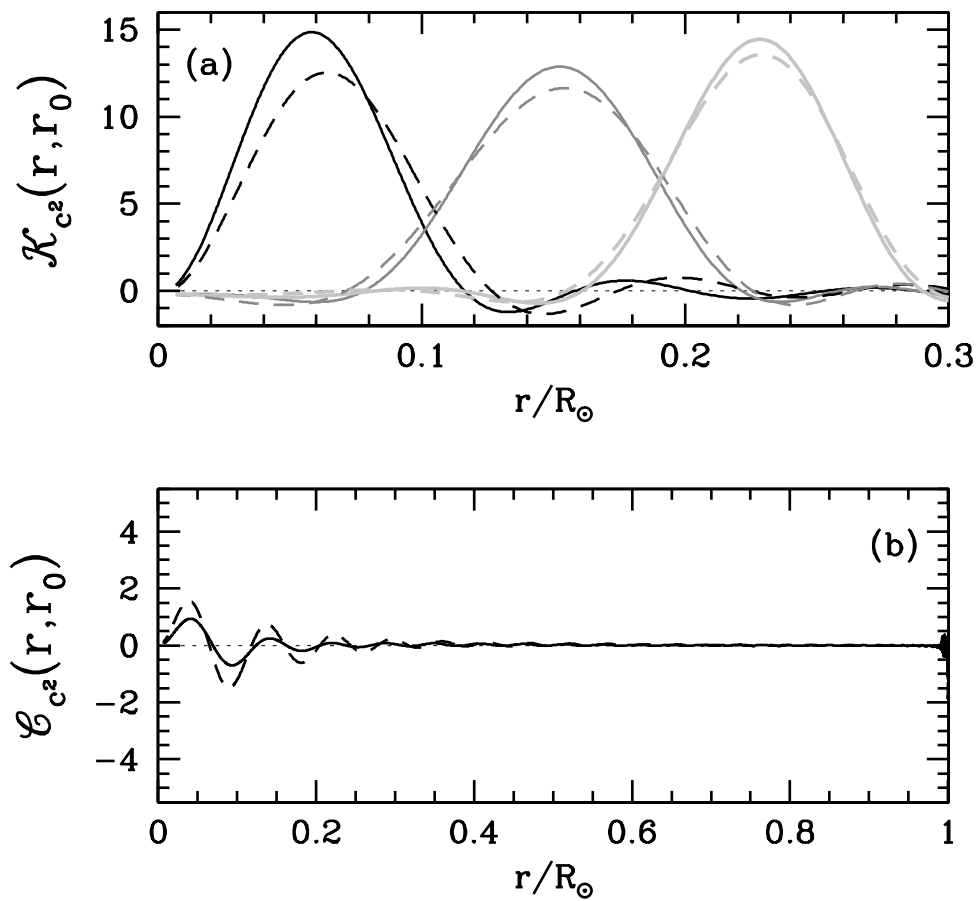}

\caption{(a) The averaging kernels obtained from the SOLA inversions
of the BiSON-13 (solid lines) and BiSON-1 sets (dashed lines). The
innermost averaging kernels and two others are shown. (b) The
cross-term kernels corresponding to the innermost averaging kernels
for the BiSON-13 (solid) and BiSON-1 (dashed) sets.  }

\label{fig:avkcsq}
\end{figure}

\begin{figure}
\epsscale{0.7}
\plotone{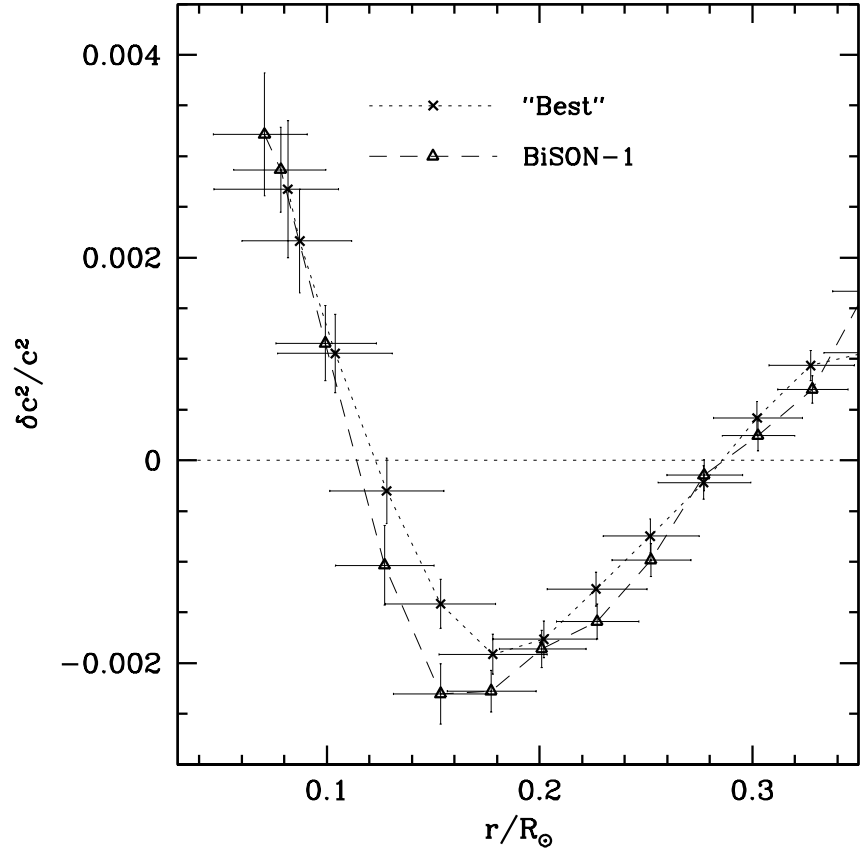}

\caption{The relative difference in the squared sound-speed between
the Sun and reference model BP04 obtained by inverting
the ``Best set''. SOLA inversion results are shown. For comparison, we also show results from set
BiSON-1.}

\label{fig:csqbest}
\end{figure}

\begin{figure*}
\epsscale{1.0}
\plotone{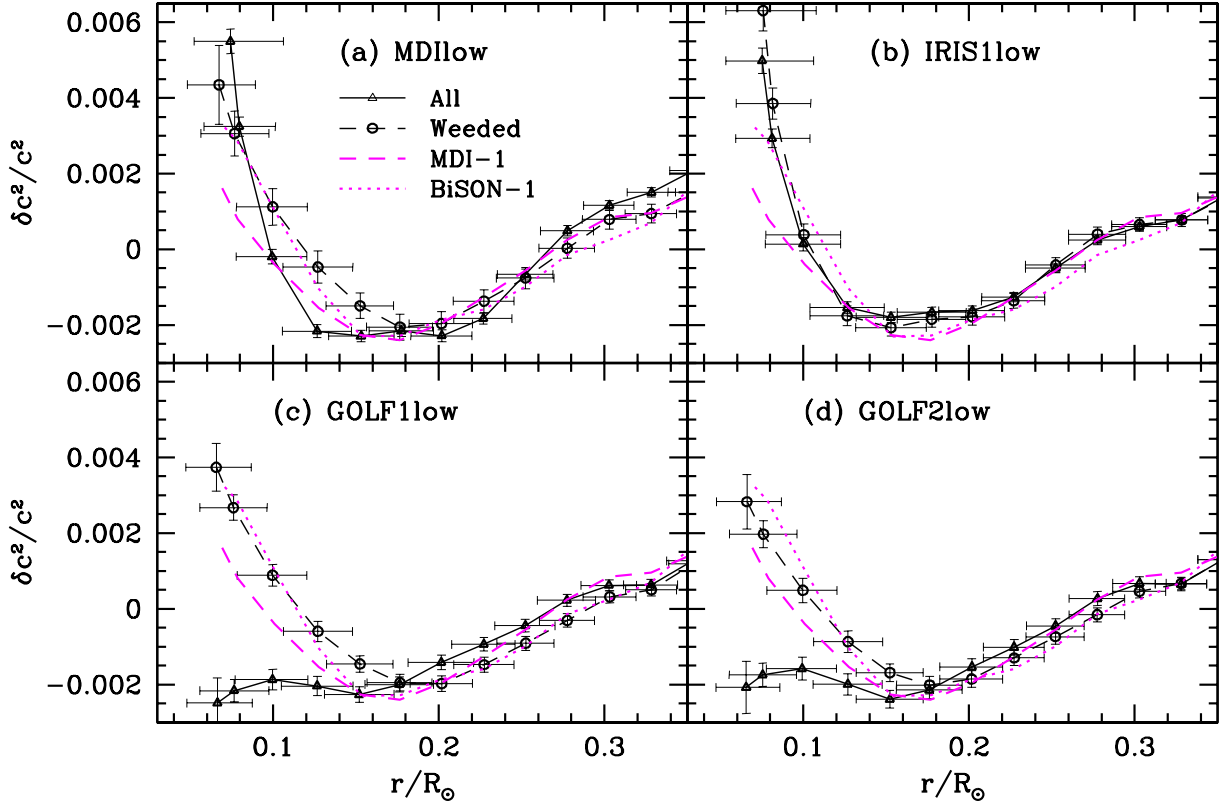}

\caption{The relative difference in the squared sound-speed between
the Sun and reference model BP04 obtained by inverting the low-degree
modes obtained with other instruments. The results for MDI-1 and
BiSON-1 are shown for comparison.  We show two results for each
external set: the ``All'' results are obtained using all the modes,
the ``Weeded'' results are obtained after weeding out modes with large
residuals. Since the same set of $\ell > 3$ modes are used for all
sets, we only focus on the core. Only SOLA inversion results are shown}

\label{fig:csq_comp}
\end{figure*}

\begin{figure}
\epsscale{0.60}
\plotone{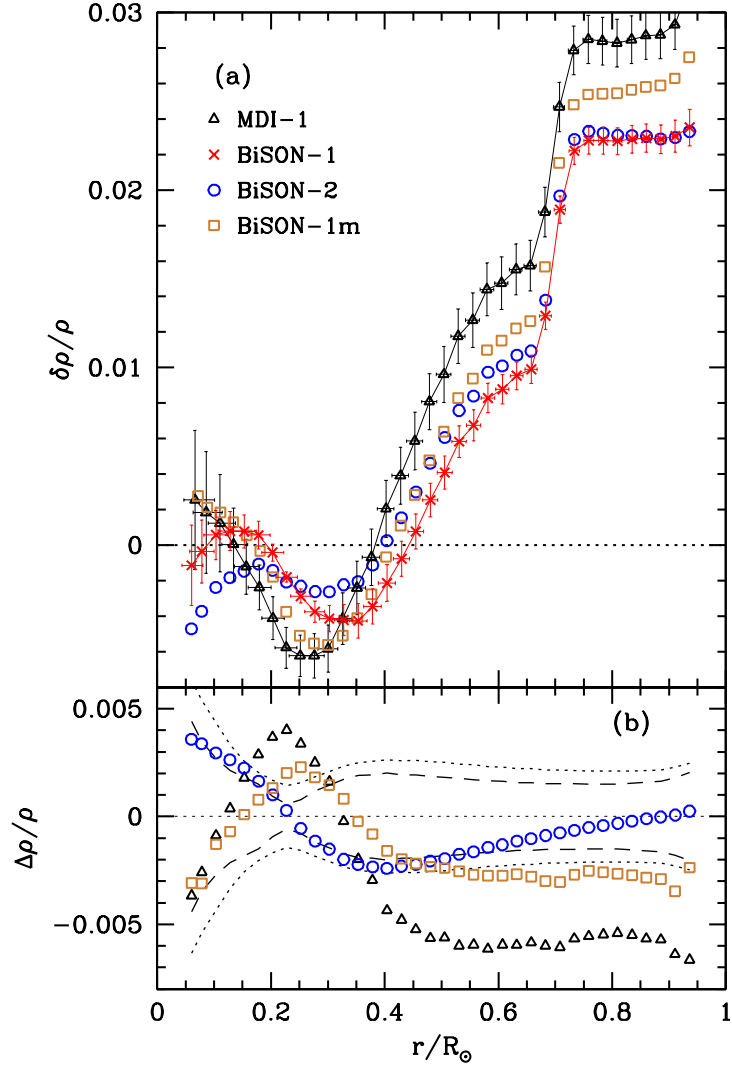}

\caption{Panel (a): Relative density differences between the Sun and
reference model BP04 obtained by inverting the data sets marked in the
figure. Only two sets of error bars are shown for the sake of
clarity. The errors on the other points are similar to those on the
BiSON-1 set. { Panel (b): The relative differences in the inferred
solar density obtained from the different data sets. The differences
are taken with respect to the solar density inferred by using the
BiSON-1 set}. The dotted line shows the 1$\sigma$ error limit for the
MDI-1 set, the dashed line is for the other BiSON sets. Note that
there are significant differences in the results. Only SOLA inversion
results are shown for the sake of clarity.}

\label{fig:bisondensity}
\end{figure}

\begin{figure}
\epsscale{0.65}
\plotone{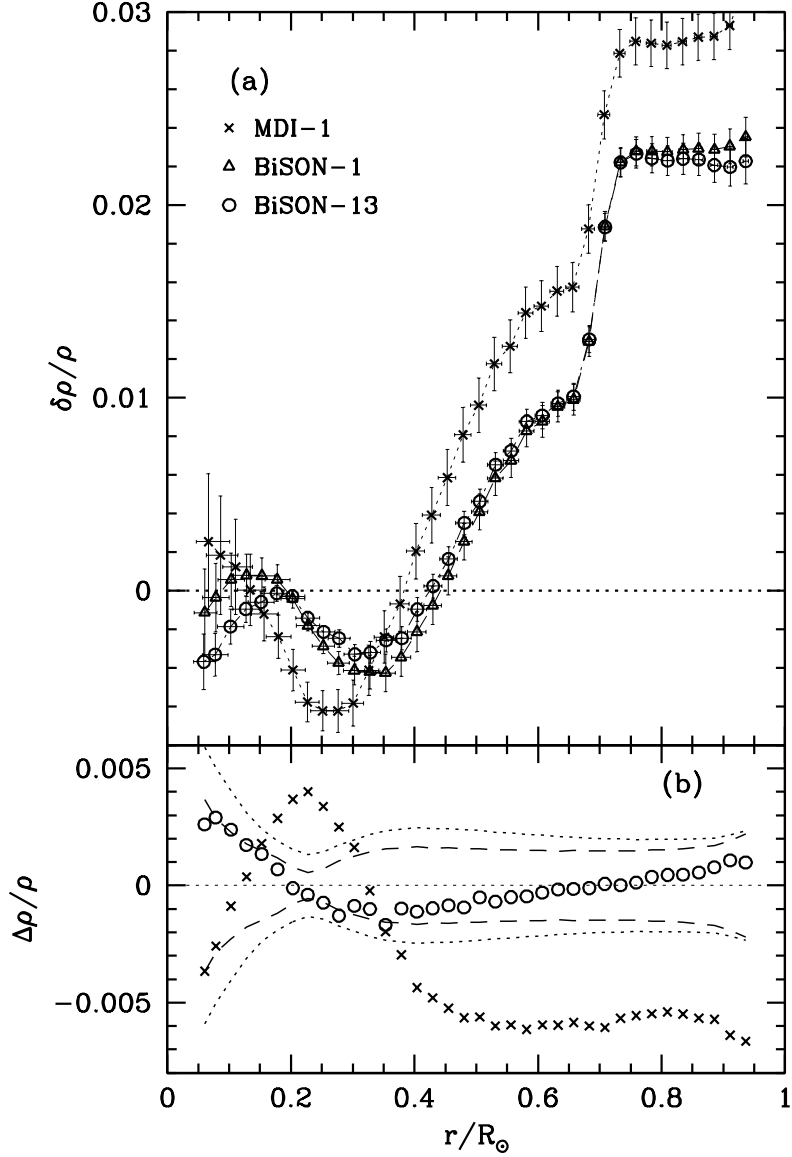}

\caption{Panel (a): Relative density differences between the Sun and
reference model BP04 obtained by inverting set BiSON-13. Results for
MDI-1 and BiSON-1 are shown for comparison. Panel (b): The relative
differences in the inferred solar sound speed obtained using the
BiSON-1 set and the BiSON-13 set. Results for MDI-1 are shown for
comparison. The dotted line shows the 1$\sigma$ error limit for the
MDI-1 set, the dashed line is for the BiSON-13 set. Only SOLA inversion results are
shown.}

\label{fig:rho13}
\end{figure}

\begin{figure}
\epsscale{0.7}
\plotone{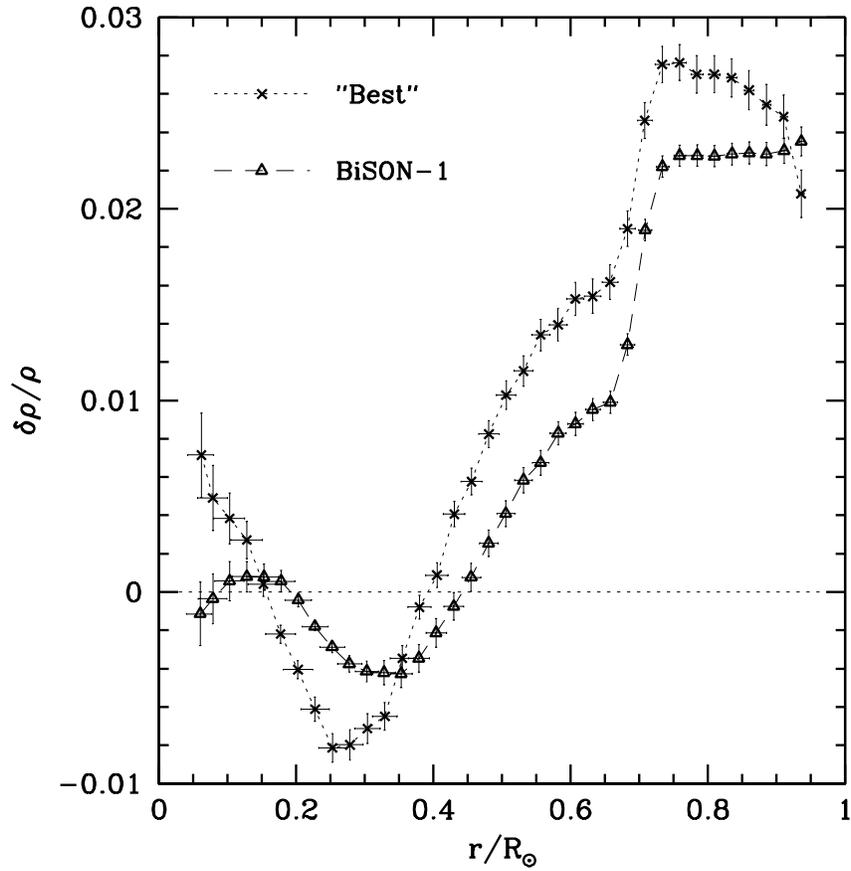}

\caption{The relative difference in the density between the Sun and
reference model BP04 obtained by inverting the ``Best set''. The
results obtained from BiSON-1 are shown for comparison. Only SOLA inversion results are shown.}

\label{fig:rho_best}
\end{figure}

The sound-speed results are very similar to what has been obtained
before, except that we go slightly deeper in the core and the errors
are slightly smaller. From Figure~\ref{fig:csqbison} we can still see
a discrepancy in the sound-speed difference around $0.2\Rsun$, a large
positive difference just below the base of the convection zone, and a
gentle fall in the convection zone itself.  Of these features, the
discrepancy at the base of the convection zone is easiest to
interpret.  This feature is usually taken as evidence for mixing in
the Sun below the base of the convection zone, mixing that is absent
in the models (see e.g., Gough et al. 1996; Basu et al. 1996. 1997;
etc.). This supposition is supported by inversions for the helium
profile of the Sun (Antia \& Chitre 1997). Support for this
interpretation also comes from the fact that models that incorporate
mixing below the base of the convection zone show a reduced
discrepancy with respect to the Sun (see e.g., Gough et al. 1996; Basu
et al. 2000b).  There is also a mismatch in the position of the
convection zone base in the Sun ($0.713\Rsun$; Christensen-Dalsgaard
et al. 1991; Basu 1998) and our reference model
($0.7146\Rsun$), and this also contributes to the discrepancy.

The discrepancy close to the core is more difficult to interpret, but
is believed to be related to weak mixing at some time during the Sun's
evolution (see e.g. Gough et al. 1996).  The third apparent
discrepancy is for $r > 0.8R_\odot$. Basu et al. (2003) noted that for
their reference model, the discrepancy decayed with depth from the
surface at a rate roughly proportional to $c^{-1}$. Since they could
not find any physical reason for such a discrepancy in what is
essentially an adiabatically stratified region, they believed that the
feature was a systematic error in the inversions caused by imperfect
suppression of the near-surface errors.

Before we try to interpret the features in the sound-speed difference
between the Sun and our models, it is instructive to look at
differences between different standard solar models and the
differences these models show with respect to the Sun.
Figure~\ref{fig:modelfig} shows the relative sound-speed and density
differences between model S and BP04, and between model BSB and
BP04. These are exact differences, but convolved with the averaging
kernels obtained from inverting set BiSON-1. Convolution with the
averaging kernels makes it easier to compare these differences with
those obtained from the inversions using the BiSON data.
Figures~\ref{fig:csqSa} and \ref{fig:csqSb}\ show the sound speed
differences, respectively for all radii and just for the core ---
between the Sun and models S and BSB; while Figure~\ref{fig:rhos}
shows the density differences between these models and the Sun.  The
first thing to note from Figure~\ref{fig:modelfig}(a) is that the
sound-speed difference between BSB and BP04 has the same type of
behavior at $r > 0.8R_\odot$ as does the sound-speed difference
between the Sun and model BP04 (and model S), and in fact the
sound-speed differences between the Sun and model BSB are flatter in
this region (Figure~\ref{fig:csqSa}).  Yet the model differences are
exact and therefore should not suffer from imperfect suppression of
near-surface errors. There is a small difference between the convolved
and the unconvolved differences that points to effects of the finite
resolution of the averaging kernels, but the effect of finite
resolution is to reduce the sound-speed difference for $r >
0.8R_\odot$.  The difference between the sound-speed profiles of BSB
and BP04 in the convection zone is a result of using different
low-temperature opacities (Ferguson et al. 2005 in BSB and Alexander
\& Ferguson 2004 in BP04). Although, the bulk of the convection zone
is stratified adiabatically, the structure closer to the surface is
affected by differences in low-temperature opacities, which is
manifested as the observed sound-speed differences in the convection
zone of these models. This occurs predominantly because of differences
in the mixing length parameter needed to model the present-day Sun. In
particular, we find that the mixing length parameter is 2.19 when OPAL
opacities are used in conjunction with the low-temperature opacities
of Ferguson et al. (2005); is 2.21 when OPAL opacities are replaced by
OP opacities; and is 2.11 when we use a combination of OP opacities
and the Alexander \& Ferguson (1994) low-temperature opacities.
Therefore, we conclude that the differences in sound-speed between the
Sun and our models in the convection are most likely to be due to
deficiencies of our models and not merely due to limitations of our
inversion techniques.

The next prominent feature in Figure~\ref{fig:modelfig}(a) is the
increased difference just below the base of the convection zone. While
at first glance this is reminiscent of the increase in
Figures~\ref{fig:csqbison} and \ref{fig:csqSa}, a careful look shows
that the difference actually occurs over a wider radius range, and is
also smaller than the difference between the Sun and BP04. The
differences between the models can basically be attributed to
differences in opacity. In the case of model S and BP04, both the
convection-zone heavy element abundance and the opacity tables used
are different. BP04 effectively has lower opacity (and hence a
shallower convection zone) than model S. While BP04 and BSB have the
same convection-zone metallicity, BSB used OP rather than OPAL
opacities. OP opacities are marginally larger than OPAL opacities for
conditions present at the base of the convection zone and that caused
the small spike in the sound-speed difference between these two
models. These differences do not explain the larger, and more
localized, sound-speed difference between the Sun and the three
models. Mixing below the solar convection zone remains the best
explanation, and as mentioned earlier, there is other evidence for
mixing in this region (e.g., Antia \& Chitre 1997).

Going back to the sound-speed differences between the Sun and models BP04, S and BSB 
at the core, we can see the solar sound-speed differences against all
three models show a localized dip around $0.2\Rsun$, implying that the
solar sound-speed is lower than that of the models. At even smaller
radii the sound-speed in the Sun appears to be larger than the
sound-speed in the models, at least for the case of models S and
BP04. This feature, particularly in the case of model S, has been used
as evidence for possible mixing at some time in the Sun's past (see
Gough et al. 1996). While mixing in the past is a possibility, the
evidence is somewhat less compelling when models with more updated
physics are used.  For example, if we look at the sound-speed
differences between the Sun and model BSB, the large, positive
difference in the core is reduced in size. However, the deficit around
$0.2\Rsun$ remains, although it is shifted inwards and is closer to
$0.15\Rsun$.  Basu et al. (2000a) had claimed that the dip in the
sound-speed difference is independent of the reference model and hence
intrinsic to the Sun, but it appears that their result was due to the
fact their various reference models used very similar physics
inputs. The difference in the reaction rates and opacities in BSB
compared to those in models S and BP04 are evidently large enough to
shift the position of the deficit. Thus, we might expect models with
newer inputs to show other changes.  Also notable is that the sound
speed difference between BSB and BP04 (see Figure~\ref{fig:modelfig})
shows a larger sound-speed in model BSB at low radii, and a lower
sound speed around $0.3\Rsun$. These differences are purely due to
changes in input microphysics, i.e., opacities as well as nuclear
reaction rates.

\begin{figure}
\epsscale{0.7}
\plotone{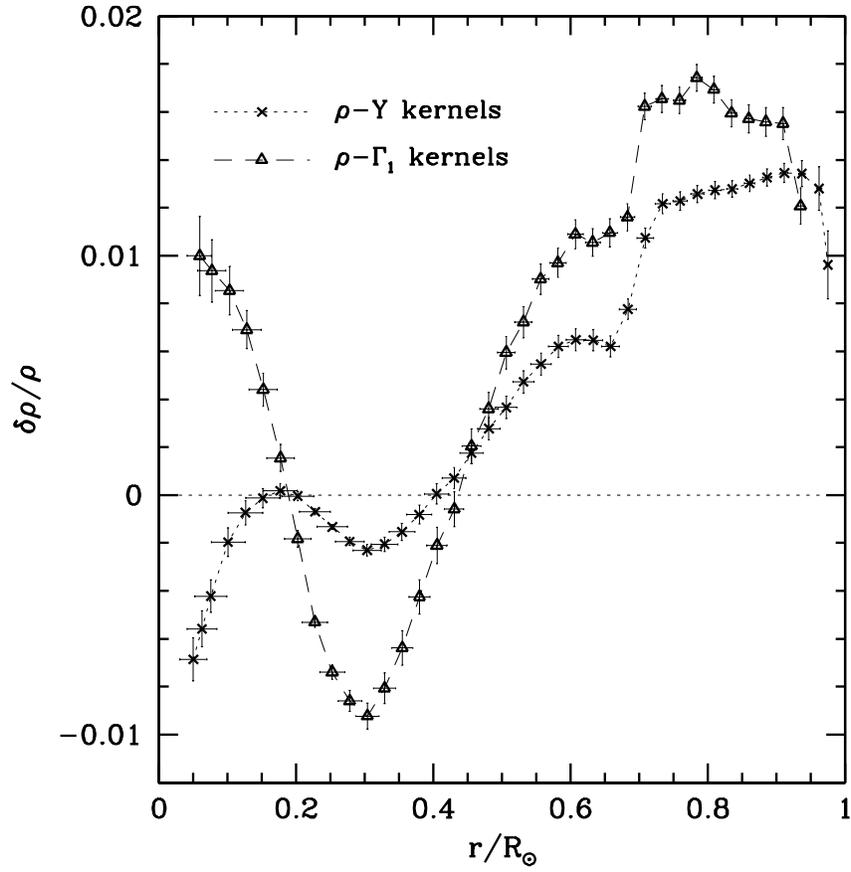}

\caption{The relative difference in the density between the Sun and
reference model S obtained by obtaining by inverting the ``Best
set''. Two results are shown, one obtained using the $(\rho,Y)$ kernel
combination, the other using the $(\rho,\Gamma_1)$ kernel
combination. Note the striking differences between the two results,
especially in the core. Only SOLA inversion results are shown.}

\label{fig:rho_yker}
\end{figure}

\begin{figure}
\epsscale{0.7}
\plotone{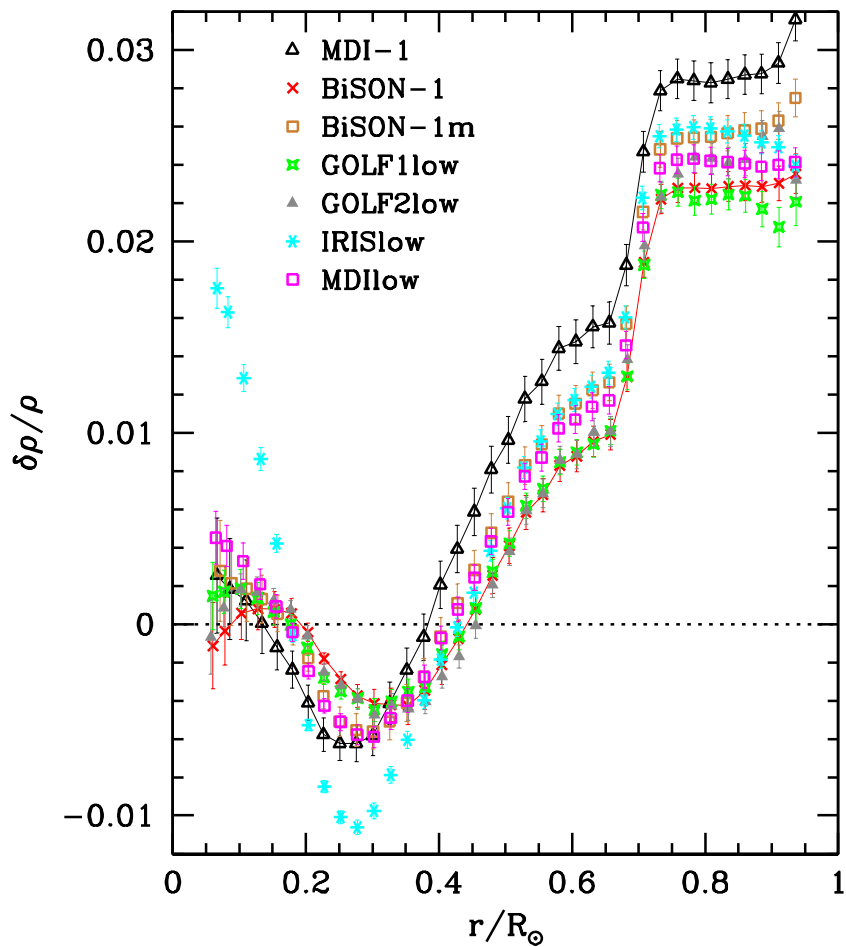}

\caption{The relative difference in the density between the Sun and
reference model BP04 obtained by inverting the low-degree modes
obtained with other instruments. The results for MDI-1 and BiSON-1 are
shown for comparison.  Only the results obtained with the weeded sets
are shown. Lines have been drawn through the BiSON-1 and MDI results
to guide the eye. Only SOLA inversion results are shown.}

\label{fig:rhoall}
\end{figure}

\begin{figure}
\epsscale{1.0}
\plotone{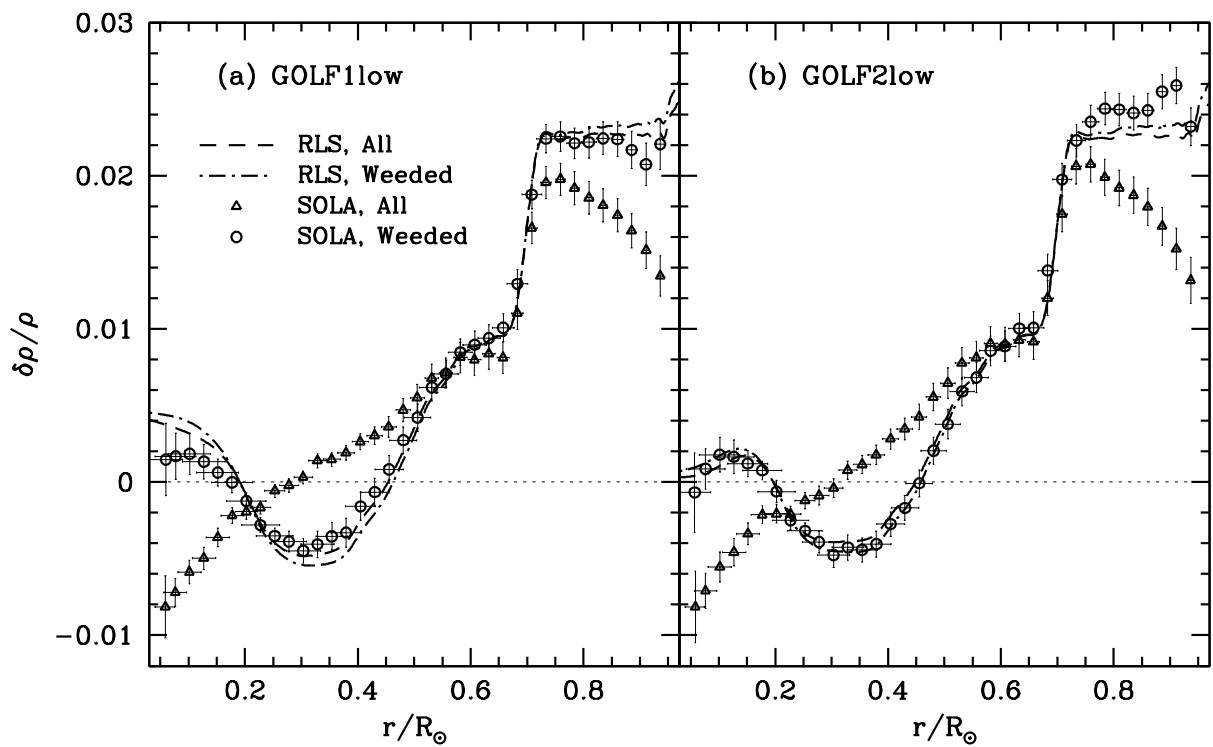}

\caption{The relative difference in the density between the Sun and
reference model BP04 obtained by inverting the two GOLF sets.  Both
RLS (lines) and SOLA (points) inversion results are shown. Note that
for the weeded sets the RLS and SOLA results match, while there is a
large difference between RLS and SOLA results when all the modes are
used. { It should be noted that the same inversion parameters were
used for both the ``Weeded'' and ``All' sets}.}

\label{fig:golf}
\end{figure}

\begin{figure*}
\epsscale{1.1}
\plottwo{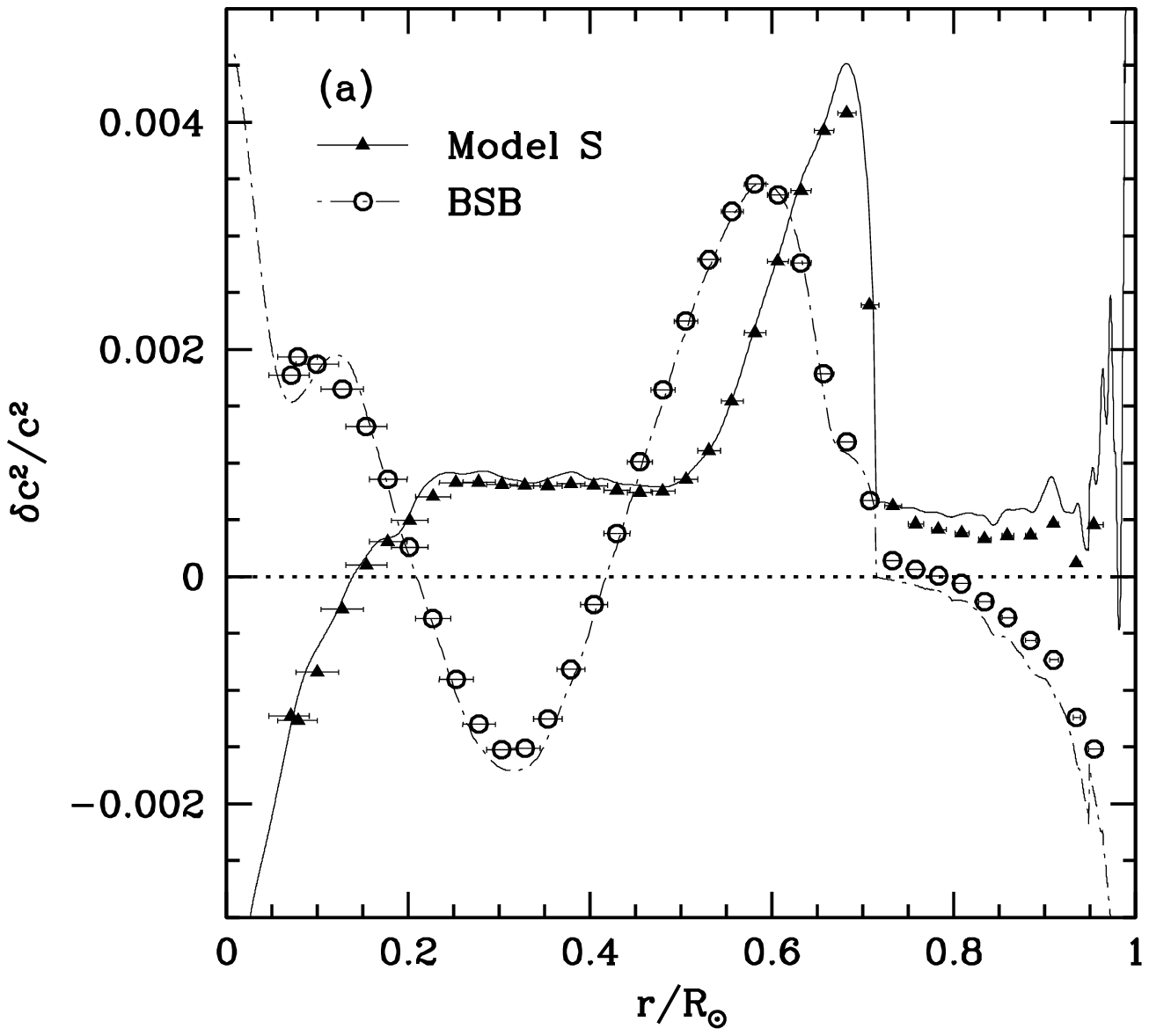}{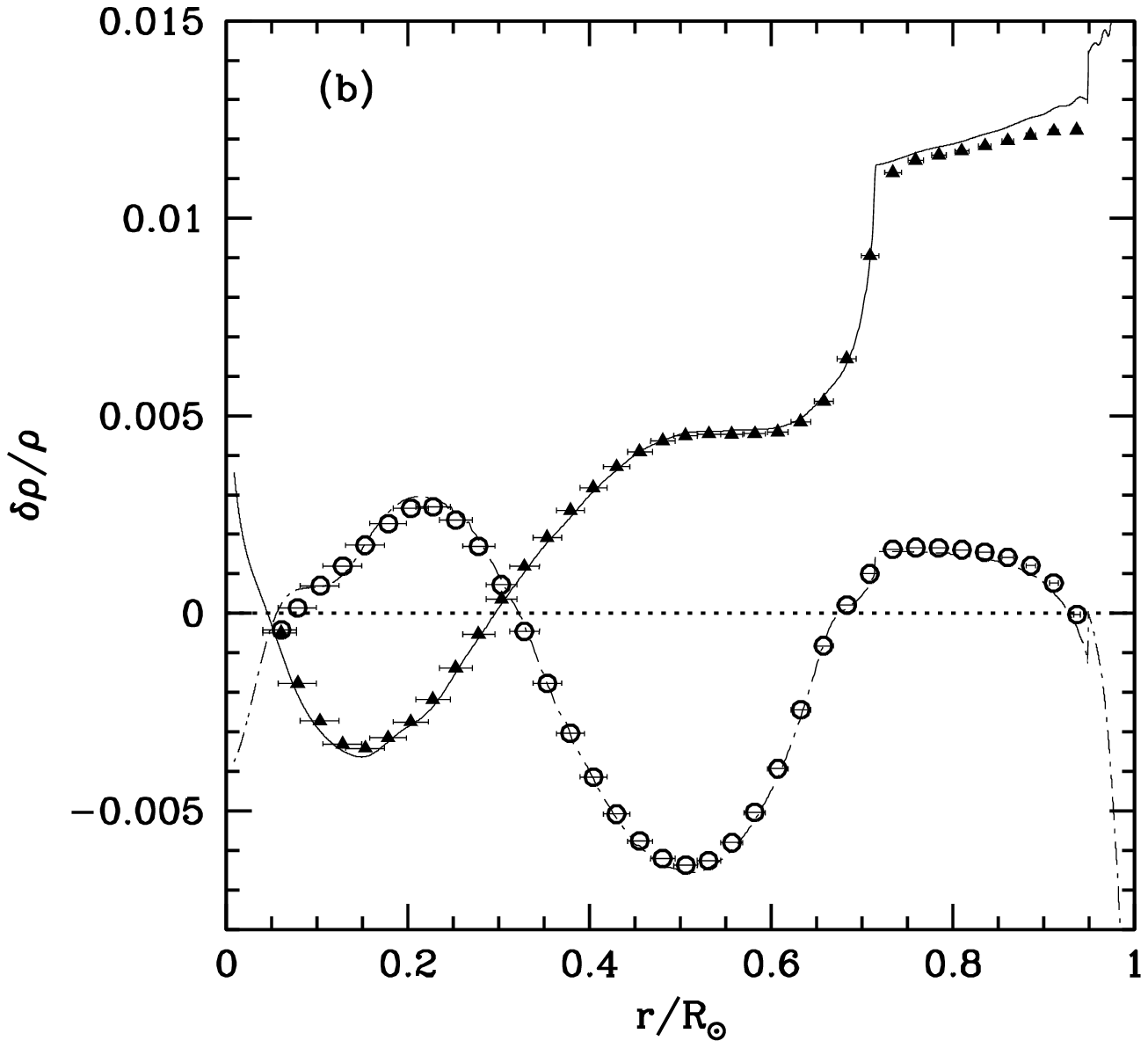}

\caption{ The relative differences in squared sound speed (panel a)
and density (Panel b), between model S and BP04 and model BSB and
BP04.  The results are in the sense S$-$BP04 and BSB$-$BP04. The lines
are the exact differences. The points are the exact differences
convolved with the averaging kernels obtained by inverting
BiSON-1. Thus, the points are what we would expect to see if we
inverted the frequency differences between the models. The differences
between the points and the lines are caused by the finite width of the
averaging kernels as well as by any non-local features that they may
have.}

\label{fig:modelfig}
\end{figure*}

\begin{figure*}
\epsscale{1.0}
\plotone{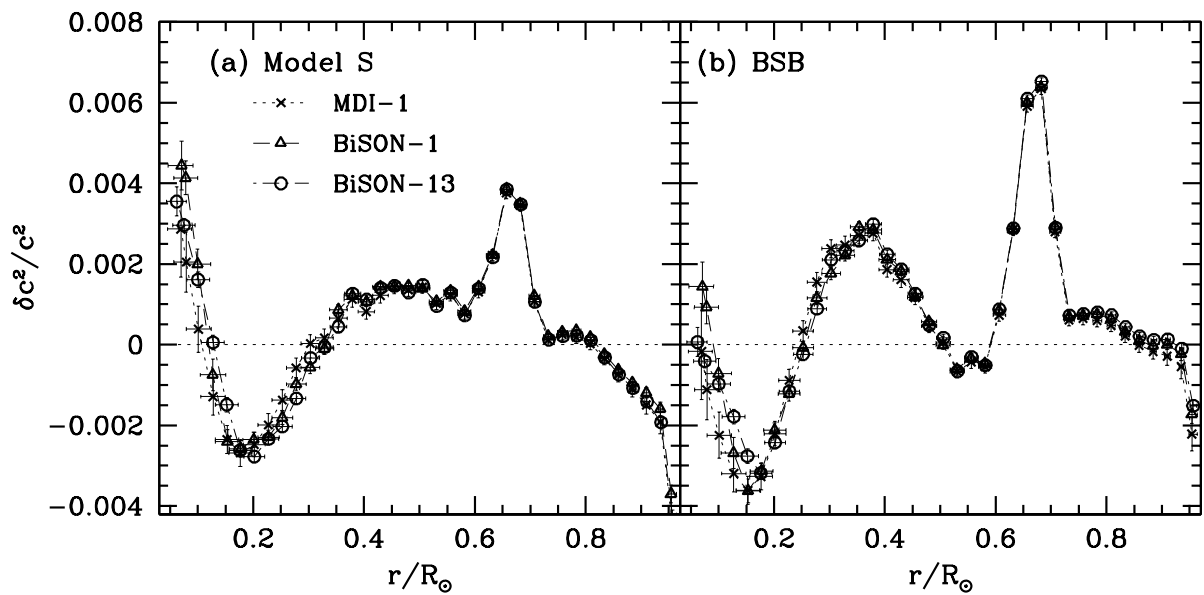}

\caption{The relative differences in squared sound speed between the
Sun and model S (panel a) and model BSB (panel b) obtained by
inverting the MDI-1, Bison-1 and BiSON-13 sets. Only SOLA inversion results are
shown.}

\label{fig:csqSa}
\end{figure*}

\begin{figure*}
\epsscale{1.0}
\plotone{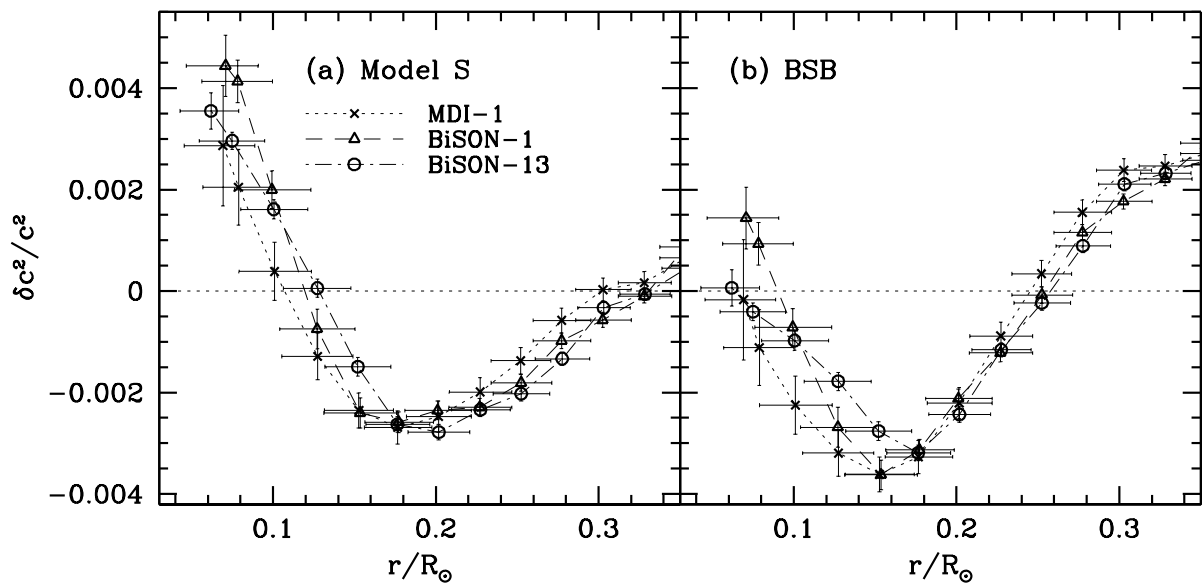}

\caption{The same as Figure~\ref{fig:csqSa}, but focusing on the solar
core. Only SOLA inversion results are shown.}

\label{fig:csqSb}
\end{figure*}

\begin{figure*}
\epsscale{1.0}
\plotone{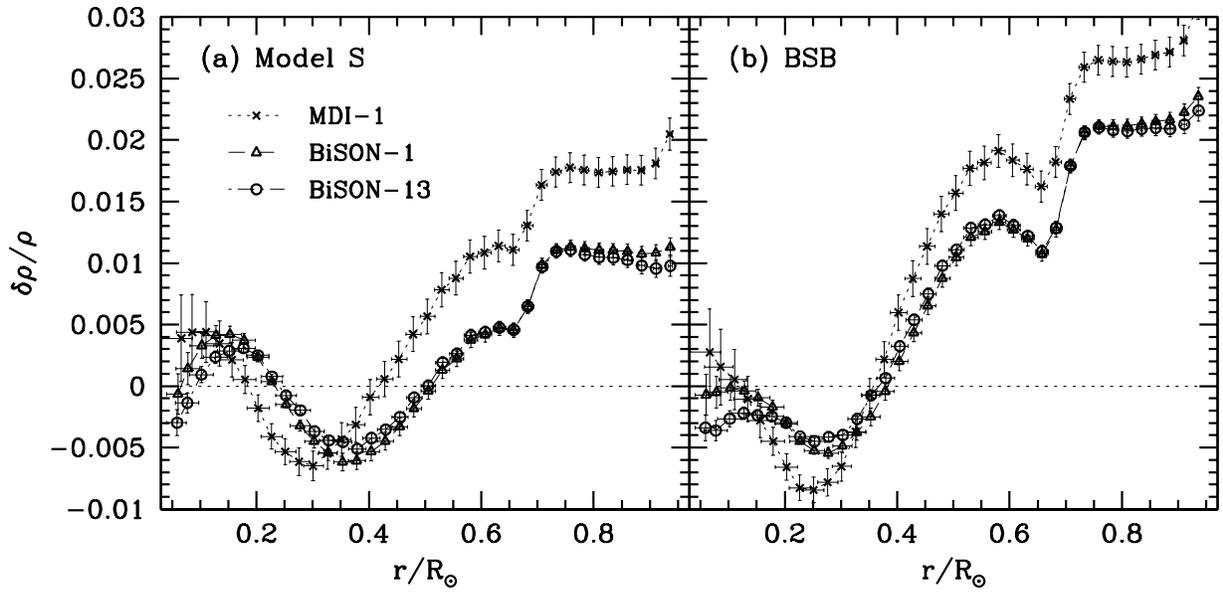}

\caption{The relative density differences between the Sun and model S
(panel a) and the Sun and model BSB (panel b), obtained from different
data sets. Only SOLA inversion results are shown.}

\label{fig:rhos}
\end{figure*}

Of course, the conjecture that there may have been mixing in the solar
core was not based on sound-speed inversions alone. Gough et
al. (1996) used the density inversion results as corroboration.  One
should note, however, that the density difference that Gough et al. (1996)
used was obtained with $(\rho, Y)$ kernels, and like the
$(\rho, Y)$ result shown in Figure~\ref{fig:rho_yker}, showed that
density of the solar core is less than the density in model S in the
core. Gough et al. (1996) argued that the relatively steep positive
gradient in $\delta\rho/\rho$ in the core and immediately beneath the
convection zone imply that the magnitude of the negative gradient of
density is too high in the model. However, our more robust inversions
show that the gradient of $\delta\rho/\rho$ is not steep for 
any of the three models, and not positive for models BP04 and BSB.

Thus, we find that even with improved data, interpreting the
sound-speed and density differences against different models is not
completely straightforward. Improvements in solar models need to go
hand-in-hand with improvements in the observational data. To be able
to invert closer to the solar core we require reliable frequency
estimates of very low-$n$, low-degree modes. Since these modes have
large amplitudes close to the core, they will help (even in the
absence of g-modes) to get a better handle on the innermost layers if
the Sun.  As we have seen in this paper, changes in the low-degree
mode set can lead to large variations in estimated density differences
between reference solar models and the Sun.  There is therefore a
clear need for very low-frequency modes if we are to obtain robust,
precise density estimates for the Sun.

\acknowledgements This paper utilizes data collected by the Birmingham
Solar-Oscillations Network (BiSON), which is funded by the UK Science
Technology and Facilities Council (STFC). We thank the members of the
BiSON team, colleagues at our host institutes, and all others, past
and present, who have been associated with BiSON.  This paper also
utilizes from the Solar Oscillations Investigation / Michelson Doppler
Imager (SOI/MDI) on SoHO.  SB would like to thank the HiROS group at
the University of Birmingham, U.K., for their hospitality during the
period when this work was conceived and developed. She would also like
to thank the Institute for Advanced Study, Princeton for its
hospitality while this paper was being written. SB acknowledges
partial support from NSF grants ATM-0348837 and ATM-0737770.


\begin{deluxetable}{llll}
\tablecolumns{4}
\tablewidth{0pc} 
\tablecaption{Frequency datasets used for the inversions}
\tablehead{\colhead{Dataset}&
           \colhead{Comments}&
           \colhead{Reference}} \startdata
          MDI-1& $0 \le l \le 150$& Schou et al. (1998)\\
     BiSON-1& 1\,yr, asymmetric fitting& \\
     BiSON-1m& Same as BiSON-1, but restricted to the \\
             &same low-$\ell$ modes as MDI-1& \\
     BiSON-2& 1\,yr, symmetric fitting& \\
BiSON-13& 13\,yr, asymmetric fitting, solar-cycle corrected& \\
        GOLF1low& $\nu \le 1800\,\rm \mu Hz$& Bertello et al. (2000)\\
        GOLF2low& $\nu \le 1800\,\rm \mu Hz$& Garcia et al. (2001), Table~V\\
         MDIlow& $\nu \le 1800\,\rm \mu Hz$& Toutain et al. (1998)\\
        IRISlow& $\nu \le 1800\,\rm \mu Hz$& Fossat et al. (2003)\\
\enddata
\label{tab:data}
\end{deluxetable}

\begin{deluxetable}{llll}
\tablecolumns{4}
\tablewidth{500pc} 
\tablecaption{Inputs and characteristics of reference solar models}
\tablehead{\colhead{\ }&
           \colhead{\ }&
           \colhead{Model}&
           \colhead{\ }}
\startdata
\small
           &     BP04                &       S                   &  BSB \\
\hline
{\bf Inputs:} \\
\\
Mass       & $1.989\times10^{33}$g   & $1.989\times10^{33}$g     & $1.989\times10^{33}$g \\
Radius     & $6.9598\times10^{10}$cm & $6.96\times10^{10}$cm     & $6.9598\times10^{10}$cm \\
Luminosity & $3.8418\times10^{33}$ erg/s & $3.846\times10^{33}$ erg/s & $3.8418\times10^{33}$ erg/s\\
\\
(Z/X)$_{\rm sur}$ &  0.0229      &   0.0245 & 0.0229\\
       &  Grevesse \& Sauval (1998) & Grevesse \& Noels (1993) & Grevesse \& Sauval (1998)\\
\\
Eq. of State &  OPAL      &       OPAL    &              OPAL\\
              &Rogers \& Nayfonov (2002)   & Rogers et al. (1996)  &  Rogers \& Nayfonov (2002)\\
\\
Opacity &   OPAL & OPAL & OP\\
        &Iglesias \& Rogers (1996)  &  Rogers \& Iglesias (1992) & Badnell et al.~(2005) \\
        & Alexander \& Ferguson (1994) & Kurucz (1991) & Ferguson et al. (2005) \\
\\
Reaction rates & Bahcall \&  &  Bahcall \&  &  Bahcall \&   \\
               & Pinsonneault (2004) & Pinsonneault (1995) & Pinsonneault (2004)\\
 & &            & Formicola et al. (2004)\\
\\
Diffusion rates &  Thoul et al. (1994) & Proffitt \&  & Thoul et al. (1994) \\
 & & Michaud(1991) \\
\\
{\bf Characteristics:} \\
\\
$Y_{\rm CZ}$ & 0.243 & 0.247 & 0.243\\
$R_{\rm CZ}$ & 0.7146 & 0.7129 & 0.7138\\
\enddata
\label{tab:models}
\end{deluxetable}


\begin{deluxetable}{cccccc}
\tablecolumns{6}
\tablecaption{Solar sound-speed and density profiles as derived from BiSON-13}
\tablehead{\colhead{$r/R_\odot$} & \colhead{$c$ (cm $s^{-1}$)} & \colhead{$\sigma_c$ (cm $s^{-1}$)} &
\colhead{$r/R_\odot$} & \colhead{$\rho$ (g cm$^{-3}$)} & \colhead{$\sigma_\rho$ (g cm$^{-3}$)}}
\startdata
   6.1949E-02 &  5.1170E+07 &  9.2178E+03 &    5.8884E-02 &  1.2078E+02 &  1.1601E-01 \\
   7.4948E-02 &  5.1169E+07 &  4.3834E+03 &    7.7225E-02 &  1.0532E+02 &  7.7585E-02 \\
   1.0052E-01 &  5.0803E+07 &  4.9108E+03 &    1.0208E-01 &  8.5937E+01 &  5.1525E-02 \\
   1.2733E-01 &  4.9858E+07 &  4.4274E+03 &    1.2706E-01 &  6.9226E+01 &  3.1670E-02 \\
   1.5208E-01 &  4.8559E+07 &  4.4193E+03 &    1.5162E-01 &  5.5510E+01 &  2.0105E-02 \\
   1.7636E-01 &  4.7033E+07 &  3.5829E+03 &    1.7709E-01 &  4.3772E+01 &  1.2053E-02 \\
   2.0184E-01 &  4.5316E+07 &  3.4470E+03 &    2.0233E-01 &  3.4221E+01 &  6.9189E-03 \\
   2.2746E-01 &  4.3569E+07 &  3.0592E+03 &    2.2730E-01 &  2.6514E+01 &  4.6064E-03 \\
   2.5253E-01 &  4.1897E+07 &  2.8858E+03 &    2.5277E-01 &  2.0234E+01 &  4.3910E-03 \\
   2.7771E-01 &  4.0301E+07 &  2.5884E+03 &    2.7803E-01 &  1.5355E+01 &  4.5718E-03 \\
   3.0320E-01 &  3.8798E+07 &  2.4506E+03 &    3.0333E-01 &  1.1570E+01 &  3.9911E-03 \\
   3.2847E-01 &  3.7395E+07 &  2.3018E+03 &    3.2879E-01 &  8.6838E+00 &  3.2633E-03 \\
   3.5370E-01 &  3.6090E+07 &  2.1119E+03 &    3.5403E-01 &  6.5346E+00 &  2.5846E-03 \\
   3.7912E-01 &  3.4866E+07 &  1.9529E+03 &    3.7946E-01 &  4.9123E+00 &  1.9819E-03 \\
   4.0439E-01 &  3.3710E+07 &  1.8638E+03 &    4.0474E-01 &  3.7160E+00 &  1.5338E-03 \\
   4.2970E-01 &  3.2629E+07 &  1.7383E+03 &    4.3006E-01 &  2.8201E+00 &  1.1410E-03 \\
   4.5504E-01 &  3.1604E+07 &  1.6411E+03 &    4.5535E-01 &  2.1521E+00 &  8.9351E-04 \\
   4.8031E-01 &  3.0634E+07 &  1.5152E+03 &    4.8063E-01 &  1.6517E+00 &  6.7778E-04 \\
   5.0566E-01 &  2.9712E+07 &  1.4385E+03 &    5.0592E-01 &  1.2737E+00 &  5.4124E-04 \\
   5.3096E-01 &  2.8822E+07 &  1.3808E+03 &    5.3121E-01 &  9.8850E-01 &  4.1931E-04 \\
   5.5630E-01 &  2.7972E+07 &  1.3146E+03 &    5.5651E-01 &  7.7054E-01 &  3.3861E-04 \\
   5.8160E-01 &  2.7130E+07 &  1.2604E+03 &    5.8181E-01 &  6.0445E-01 &  2.6438E-04 \\
   6.0693E-01 &  2.6311E+07 &  1.1919E+03 &    6.0711E-01 &  4.7634E-01 &  2.1495E-04 \\
   6.3223E-01 &  2.5483E+07 &  1.1390E+03 &    6.3243E-01 &  3.7756E-01 &  1.6995E-04 \\
   6.5754E-01 &  2.4637E+07 &  1.0619E+03 &    6.5772E-01 &  3.0095E-01 &  1.4047E-04 \\
   6.8283E-01 &  2.3706E+07 &  1.0045E+03 &    6.8304E-01 &  2.4197E-01 &  1.1334E-04 \\
   7.0811E-01 &  2.2614E+07 &  9.3680E+02 &    7.0838E-01 &  1.9685E-01 &  9.3937E-05 \\
   7.3337E-01 &  2.1257E+07 &  8.9853E+02 &    7.3371E-01 &  1.6281E-01 &  7.8683E-05 \\
   7.5869E-01 &  1.9881E+07 &  8.7096E+02 &    7.5903E-01 &  1.3311E-01 &  6.5124E-05 \\
   7.8404E-01 &  1.8495E+07 &  7.7763E+02 &    7.8435E-01 &  1.0703E-01 &  5.4001E-05 \\
   8.0934E-01 &  1.7089E+07 &  7.3027E+02 &    8.0967E-01 &  8.4315E-02 &  4.3385E-05 \\
   8.3464E-01 &  1.5649E+07 &  6.8648E+02 &    8.3499E-01 &  6.4667E-02 &  3.4486E-05 \\
   8.5995E-01 &  1.4156E+07 &  6.1451E+02 &    8.6031E-01 &  4.7800E-02 &  2.6671E-05 \\
   8.8527E-01 &  1.2582E+07 &  5.7896E+02 &    8.8564E-01 &  3.3507E-02 &  1.9704E-05 \\
   9.1060E-01 &  1.0881E+07 &  5.9297E+02 &    9.1096E-01 &  2.1657E-02 &  1.4339E-05 \\
   9.3590E-01 &  8.9771E+06 &  5.5548E+02 &    9.3628E-01 &  1.2157E-02 &  9.5495E-06 \\
   9.5661E-01 &  7.1388E+06 &  4.9286E+02 &    9.5754E-01 &  5.9835E-03 &  6.2548E-06 \\
\enddata
\label{tab:seismic}
\end{deluxetable}
\clearpage


\end{document}